\documentclass[aps,prd,a4paper,preprintnumbers,floatfix,nofootinbib,showpacs,superscriptaddress, notitlepage,11pt]{revtex4-1} 
\pdfoutput=1
\usepackage{color}
\usepackage[table]{xcolor}
\usepackage[colorlinks=true,citecolor=blue,linkcolor=blue,
breaklinks=true]{hyperref}
\usepackage{amsmath,amssymb}
\usepackage{graphicx}
\usepackage{slashed}
\usepackage[caption=false,labelformat=simple]{subfig}
\usepackage{wrapfig}
\usepackage{placeins}
\usepackage{multirow, makecell}
\usepackage{booktabs}

\usepackage{url}
\usepackage{enumitem}
\usepackage{multirow}
\usepackage{array}
\usepackage{tabulary}
\usepackage{float}

\usepackage{ulem}
\renewcommand{\emph}{\textit}

\newcolumntype{?}{!{\vrule width 1pt}}
\newcolumntype{M}[1]{>{\centering\arraybackslash}m{#1}}
\newcommand{\lag}{\mathcal L}

\def\mET{\slashed{E}_T}

\begin{document}
\title{Signals for a 2HDM with $Z'$ at the LHC}

\author{Anjan Kumar Barik}\email{anjanbarik@hri.res.in}
\affiliation{Regional Centre for Accelerator-based Particle Physics,
Harish-Chandra Research Institute,\\ A CI of Homi Bhabha National
Institute, Chhatnag Road, Jhunsi, Prayagraj 211019, India.}
\author{Santosh Kumar Rai}\email{skrai@hri.res.in}
\affiliation{Regional Centre for Accelerator-based Particle Physics,
Harish-Chandra Research Institute,\\ A CI of Homi Bhabha National
Institute, Chhatnag Road, Jhunsi, Prayagraj 211019, India.}
\author{Agnivo Sarkar}\email{agnivosarkar@hri.res.in}
\affiliation{Regional Centre for Accelerator-based Particle Physics,
Harish-Chandra Research Institute,\\ A CI of Homi Bhabha National
Institute, Chhatnag Road, Jhunsi, Prayagraj 211019, India.}
\preprint{HRI-RECAPP-2025-01}
\begin{abstract}
We consider a neutrinophilic $U(1)$ extension of the Standard Model (SM) under which only a second Higgs doublet and SM singlet scalars and fermions are charged. The new gauge boson $Z'$ couples to SM minimally, generated by $Z-Z'$ mixing. As the $Z'$ is very weakly coupled, it can mostly be produced through the decay of the scalars from the second Higgs doublet at the Large Hadron Collider (LHC). We discuss the scalar sector of the model in detail and consider decay modes such as $(H^{\pm} \to W^\pm Z', h_2 \to VV, (V = W^\pm, Z, Z'), A_2 \to h_1 Z'(Z))$ that lead to multilepton signals at the LHC from the pair production of the scalars. We analyze the signal with a representative value of the $Z'$ mass to show the discovery potential of the 2HDM scalars at the LHC. 
\end{abstract}
\maketitle
\section{Introduction}
The discovery of the $125$ GeV scalar at the Large Hadron Collider (LHC) has completed the search for the SM particle content \cite{ATLAS:2012yve,CMS:2012qbp}. Although this scalar exhibits properties consistent with the SM, it is still possible that it could belong to an extended scalar sector. Extension of the SM scalar sector by any non-trivial representation of the SM $SU(2)_L$ gauge symmetry would introduce new CP odd and charged Higgs bosons and other CP even scalars. As the SM does not contain any CP odd and charged Higgs scalars, the discovery of these scalars will point toward the BSM physics. Among the various extensions of the SM scalar sector, adding an extra $SU(2)_L$ doublet scalar has been of great interest and is known as the two-Higgs doublet model (2HDM) \cite{Branco:2011iw}. Unlike more complex multiplets, incorporating the doublet does not alter the electroweak precision observable, the $\rho$ parameter, at the tree level. The 2HDM's have been used in 
models to generate non-zero neutrino mass at tree level $(\nu\text{2HDM})$ \cite{Ma:2000cc,Gabriel:2006ns,Davidson:2009ha} or radiatively \cite{Ma:2006km}. An inert 2HDM has been popular in explaining dark matter (DM) abundance in our Universe \cite{Ma:2006km,Abdallah:2024npf} and 2HDM's have also been used to address some key issues of the SM, such as the strong CP problem and matter anti-matter asymmetry \cite{Kim:1986ax,Trodden:1998qg}.   
  
Although one can introduce additional scalar doublets to the SM particle spectrum trivially, they would also appear if the SM gauge symmetry is extended. Many well-motivated BSM scenarios, {\it e.g.} MSSM \cite{Martin:1997ns}, Type-II seesaw \cite{Konetschny:1977bn}, and LRSM \cite{Mohapatra:1974hk} incorporate such additional scalars. 
In this work, we focus on such an addition where the extra scalar 
doublet appears in a $U(1)_X$ extended scenario and helps in generating mass for the light neutrinos as well as provide a DM candidate \cite{Abdallah:2021dul}.  
All the SM fields are singlets under the new gauge symmetry and therefore couple to the new gauge boson $Z'$ through its mixing with the SM $Z$ boson.  
The scalar sector consists of an additional $SU(2)$ Higgs doublet and two SM singlet scalars and we also include three generations of SM singlet vector-like fermions. As the $Z'$ has very weak couplings to the SM fermions, the experimental constraints on the $Z'$ mass allow it to be much lighter than 
the exotic scalars that can be produced directly at the LHC. These scalars can then decay to the $Z'$ and lead to interesting signals, which we analyze in this work.

The LHC has carried out searches for both BSM CP-even and CP-odd scalars, where it is produced directly or associatively with the SM particles \cite{ATLAS:2015iie,ATLAS:2017ayi,CMS:2018amk,CMS:2019pzc,CMS:2019qcx,ATLAS:2019tpq}.  
The LHC has mostly focused on searches of a charged Higgs boson where it is singly produced in association with top and bottom quarks \cite{ATLAS:2018gfm,ATLAS:2018ntn}. Surprisingly, there are no dedicated searches for pair-produced charged Higgs by either ATLAS or CMS Collaborations. In this work, we consider the pair production of all combinations of the scalars belonging to the second Higgs doublet at the LHC, 
which then decay to gauge bosons $(H^{\pm} \to W^\pm Z', h_2 \to V V, (V = W^\pm, Z, Z'), A_2 \to h_1 Z' (Z)  )$. 
The subsequent decay of the gauge bosons can produce interesting multilepton and multi-jet signals. We choose an inclusive $4\ell +X$ final state to look for at the LHC. We also 
evaluate the current constraints on our model parameters using the multilepton signal searches at ATLAS and CMS \cite{ATLAS:2021kog,CMS:2017moi}. 

This model contains BSM fermions and scalars field which help generate SM neutrino masses via the inverse seesaw mechanism and address the DM puzzle of the Universe \cite{Abdallah:2021npg,Abdallah:2024npf}. The collider signals of the new gauge boson $Z'$ in this model have been studied at the LHC, where the new physics signal is mainly driven by gauge kinetic mixing (GKM) \cite{Abdallah:2021npg}. Additionally, we have also studied the scenario where the GKM is very small, and the $Z'$ production is driven by the SM Higgs boson production (considering scalar mixings) that decays to the $Z'$ pair \cite{Abdallah:2021dul}. The signature for an invisible $Z'$ that decays dominantly to DM particles was proposed at the future muon collider in $\gamma + \mET$ final state utilizing the radiative return feature \cite{Barik:2024kwv}. In this work, we focus on the 2HDM sector of the model 
that can have interesting and hitherto unexplored signals when the $Z'$ is a byproduct of its decay. 

The paper is organized as follows. In section \ref{sec:model} we briefly describe the model and its particle spectrum. In section \ref{sec:scalar search} we discuss how the existing multilepton searches constrain our parameter space. In section \ref{sec: collider} we present collider analysis in $4 \ell + X$ final state at LHC with $14$ TeV center of mass energy. Finally, we summarize and conclude in section \ref{sec:summary}. 

\section{The Model}
\label{sec:model}
We consider a BSM model where the SM gauge symmetry is extended with an additional $U(1)_{X}$ gauge symmetry. Apart from the SM particle content the model contains two SM singlet scalars $(S, S_{2})$, three generations of two chiral fermions $(N^i_L, N^i_R)$ which are SM gauge singlet, and an additional  SM $SU(2)$ doublet scalar $(H_{2})$. The charge assignment of these new particles along with the SM-like scalar doublet $H_1$ are presented in Table.\ref{tab:particle}. 
\begin{table}[!h]
\begin{center}
\begin{tabular}{|c|c|c|}
\hline 
Fields  & $SU(3)_C \times SU(2)_L \times U(1)_Y \times U(1)_X$ & Spin \\
\hline 
$N_L^i$ & $\left(1\,,  1 \,, 0 \,, 1 \right)$ & 1/2 \\ [1mm]
\hline 
$N_R^i$ & $\left(1\,,  1 \,, 0 \,, 1\right)$ & 1/2 \\ [1mm]
\hline
$H_1$  & $\left(1 \,, 2 \,, -\frac{1}{2} \,, 0 \right)$ & 0 \\ [1mm]
\hline 
$H_2$ & $\left(1 \,, 2 \,, -\frac{1}{2}\,,  -1 \,\right)$ & 0 \\ [1mm]
\hline 
$S$ & $\left(1\,,  1 \,, 0\,,  2\right)$ & 0  \\ [1mm] 
\hline 
$S_2$ & $\left(1\,,  1\,,  0\,, -1\right)$ & 0 \\ [1mm]
\hline
\end{tabular}
\end{center}
\caption{New fields and their charge assignments under the SM gauge
group and $U(1)_X$.}
\label{tab:particle}
\end{table}
The SM fermions do not transform under the new abelian gauge symmetry and are inert to the new $U(1)$. 
The Lagrangian containing  all the BSM fields and $H_1$ is given by
\begin{eqnarray}
  \lag_{BSM+H_1} =  \lag_{\rm{vector}} +  \lag_{\rm{fermion}} +  \lag_{\rm{scalar}}
\end{eqnarray}
The covariant derivatives for the fields $H_1,H_2,S,S_2$ and $N_{L/R}$ are defined as
\begin{align}
D_\mu^{(H_1)} &= \partial_\mu -ig_2 \frac{\sigma^a}{2}W_\mu^a + i\frac{g_1}{2}B_\mu  \, ,\nonumber \\
D_\mu^{(H_2)} &= \partial_\mu -ig_2\frac{\sigma^a}{2}W_\mu^a +i\frac{g_1}{2}B_\mu + i g_x C_\mu  \, , \,\,\,\, D_\mu^{(S)} = \partial_\mu -2ig_x  C_\mu  \nonumber \\
D_\mu^{(S_2)} &= \partial_\mu + ig_x  C_\mu  \, , \,\,\,\, 
D_\mu^{(N_{L/R})} = \partial_\mu -ig_x C_\mu  \, , 
\end{align}
where  $B_\mu$, $W_\mu^a$, and $C_\mu$ are the gauge fields of the $SU(2)_L, U(1)_Y\, \text{and} \, U(1)_X$ gauge symmetry respectively with their corresponding gauge couplings given by $g_1$, $g_2$, and $g_x$. The kinetic part of the gauge fields in this model is given by 

\begin{eqnarray}
	\lag \supset - \frac{1}{4} G^{a,\mu\nu} G_{\mu\nu}^a - \frac{1}{4} W^{b,\mu\nu} W_{\mu\nu}^b -\frac{1}{4} B^{\mu\nu} B_{\mu\nu} - \frac{1}{4} C^{\mu\nu} C_{\mu\nu} + \frac{1}{2} \tilde{g} B^{\mu\nu} C_{\mu\nu} \, , 
    \label{Eq:gaugekin}
\end{eqnarray}
where $G^{a}_{\mu\nu}, W^{b}_{\mu\nu}, B_{\mu\nu}\, \rm{and} \, C_{\mu\nu}$ are the field strength tensor of $SU(3)_C, SU(2)_L, U(1)_Y\, \text{and} \, U(1)_X$ gauge bosons respectively. The term $\frac{1}{2} \tilde{g} B^{\mu\nu} C_{\mu\nu}$ in Eq.(\ref{Eq:gaugekin}) represents the gauge kinetic mixing between the $U(1)_Y$ and $U(1)_X$ gauge fields. After the spontaneous breaking of gauge symmetry, the gauge boson corresponding to the broken generator of the gauge group becomes massive by absorbing the goldstone mode of the scalar. The $SU(2) \times U(1)_Y$ gauge symmetry is broken to $U(1)_{\text{em}} $ when the scalars \(H_1\) and \(H_2\) obtain vacuum expectation value (vev), while the \(U(1)_X\) gauge symmetry is broken when all BSM $(H_2,S,S_2)$ scalars acquire their vevs. The vevs of the scalars \(H_1\), \(H_2\), \(S\), and \(S_2\) are denoted by \(v_1\), \(v_2\), \(v_s\), and \(v_{s_2}\), respectively. The masses for the gauge bosons are derived from the kinetic term of the scalars given by 
\begin{equation}
  \lag \supset   \left(D^\mu H_1 \right)^\dagger D_\mu H_1 + \left(D^\mu H_2 \right)^\dagger D_\mu H_2 + \left(D^\mu S \right)^\dagger D_\mu S +\left(D_\mu {S_{2}} \right)^\dagger D_\mu {S_{2}}.
\end{equation} 
This model has three neutral gauge bosons. After diagonalizing their $3\times3$ mass matrix, we find one of them is massless, which is the photon. The mass of the other two physical neutral gauge bosons $Z$ and $Z'$ is given by \cite{Abdallah:2021npg}
\begin{eqnarray}
M_{Z,Z'}^2 &=& \frac{1}{8}\Big[g_z^2 v^2 + {g'_x}^2 v^2 + 4 g_x g'_x v_2^2 + 4 g_x^2(v_2^2+ 4 v_s^2+v_{s_2}^2)\Big] \\
& &  \mp  \frac{1}{8}\sqrt{\Big({g'_x}^2 v^2 + 4 g_x g'_x v_2^2 + 4 g_x^2(v_2^2+ 4 v_s^2+v_{s_2}^2) - g_z^2 v^2\Big)^2 + 4 g_z^2 \Big(g'_x v^2 + 2 g_x v_2^2\Big)^2}  \nonumber \, , 
\end{eqnarray}
Where $g_z=\sqrt{g_1^2+g_2^2}$, and
$g'_x=\frac{g_1\tilde{g}}{\sqrt{1-\tilde{g}^2}}$. The mixing angle $\theta'$
between $Z$ and $Z'$ is given by
\begin{equation}\label{tanthetap}
\tan 2 \theta' = \dfrac{2g_z \left(g'_x v^2 + 2g_x v_2^2\right)}{{g'_x}^2 v^2 + 4 g_x g'_x v_2^2 + 4 g_x^2(v_2^2+ 4 v_s^2+v_{s_2}^2) - g_z^2 v^2}.
\end{equation}

\begin{figure}[h!]
\begin{center}
\includegraphics[width=0.49\textwidth]{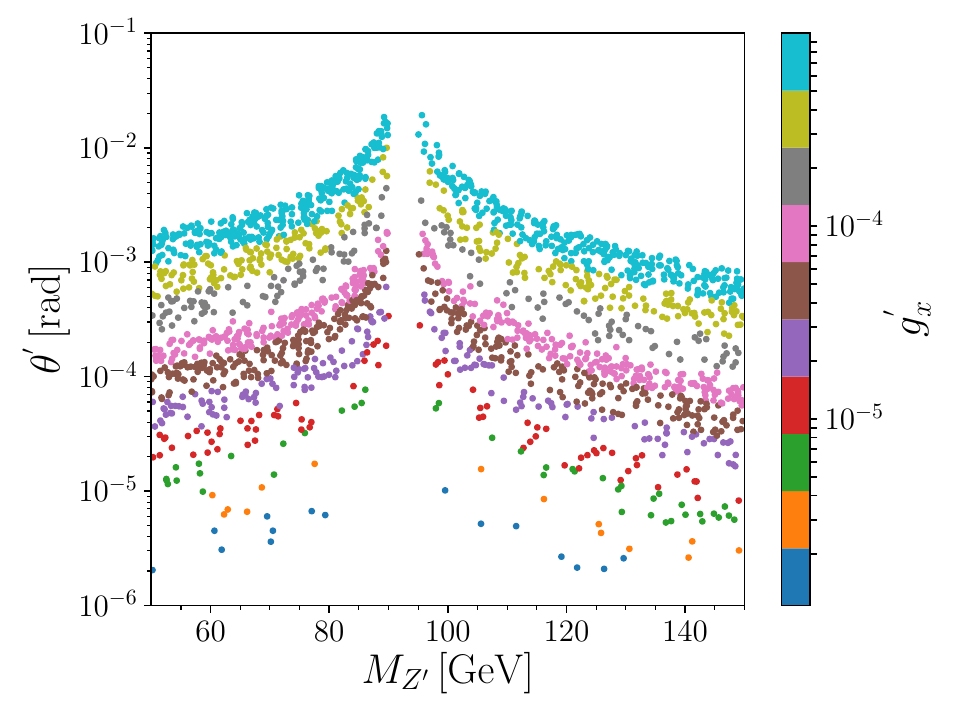}
\includegraphics[width=0.49\textwidth]{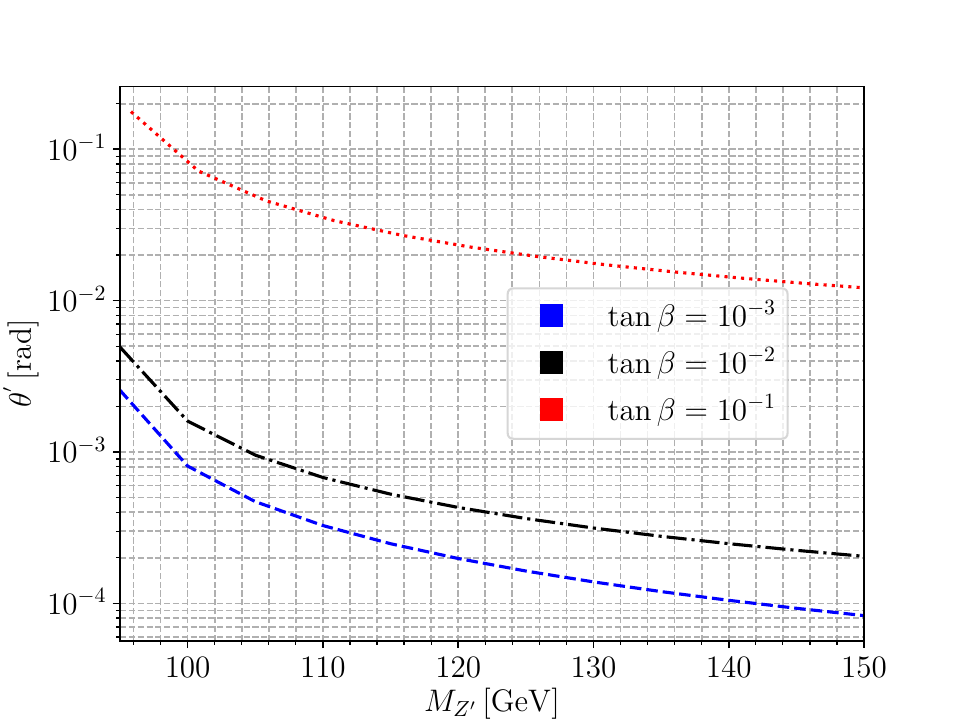}
\end{center}
\vspace{-0.5cm}
\caption{Variation of the $Z-Z'$ mixing angle $\theta'$ on $M_{Z'}$ and the gauge kinetic mixing parameter $g'_x$ (left plot). The value $\theta'$ for three values of $\tan\beta$ with $M_{Z'} \in [95,150]$ (right plot). }
\label{Fig-mzp-twp-gxp}
\end{figure}
The mixing between the $Z$ and $Z'$ bosons can alter the $Z$ boson mass and its couplings to SM fermions. Since properties of the $Z$ boson, such as its mass, decay width, and branching fractions to SM modes, have been very precisely measured at the Stanford Linear Collider (SLC), and the Large Electron-Positron (LEP) collider, and these measurements impose stringent constraints on the $Z-Z'$ mixing angle $\theta'$ to be $\leq 10^{-3}$ \cite{ParticleDataGroup:2020ssz}. 
In Fig.~\ref{Fig-mzp-twp-gxp}, we illustrate the dependence of $\theta'$ on the mass of the $Z'$ boson ($M_{Z'}$) and the redefined GKM parameter $g'_x$,  where we have set $\tan \beta = v_2/v_1 = 10^{-4}$, $v_s = 10$ GeV, $v_{s_2} = 97.98$ GeV, $M_{Z'} \in [50, 150]$ GeV, and $g'_x \in [10^{-6}, 10^{-3}]$ to generate the distribution. In the right panel of Fig.~\ref{Fig-mzp-twp-gxp}, we plot $\theta'$ for three values of $\tan \beta = 10^{-1}, 10^{-2},$ and $10^{-3}$ with $M_{Z'} \in [95, 150]$ GeV and $g'_x = 10^{-4}$. 
It is clear from the figure that $\tan \beta \sim \mathcal{O}(10^{-2})$ has an upper limit to satisfy the $Z-Z'$ mixing angle constraint. As evident from the figure and the expression for $\tan 2\theta'$, the mixing angle $\theta'$ increases with increasing $\tan \beta$. Consequently, for a light $Z'$ boson, the constraints require $g'_x \lesssim \mathcal{O}(10^{-4})$ and $\tan \beta$ has an upper limit ($\lesssim \mathcal{O}(10^{-2})$) to remain consistent with the $\theta'$ limits. 

The Lagrangian terms containing kinetic and interaction of the BSM fermions are given by
\begin{equation}
 \lag_{\rm{fermion}} = i\,\overline N_L \gamma^\mu D_\mu N_L + i\,\overline N_R \gamma^\mu D_\mu N_R-\ Y_\nu\,\overline{l_L} H_2 N_R - Y_R S \overline N_R N_R^C - Y_L S \overline N_L N_L^C - \hat{M}_N\overline{N}_L N_R + \text{h.c.}  
\end{equation}
After symmetry breaking, the SM neutrinos become massive in this model via the inverse seesaw mechanism \cite{Abdallah:2021npg}.   
The mass terms for the neutrinos are given by
\begin{eqnarray}
\lag_\nu^\text{mass} = -\frac{v_2}{\sqrt 2}Y_\nu \overline{\nu}_L N_R - \frac{v_s}{\sqrt 2}Y_R\overline{N_R^C} N_R - \hat{M}_N \overline{N}_L N_R - \frac{v_s}{\sqrt 2}Y_L\overline{N_L^C} N_L + {\rm h.c.}
\end{eqnarray}
 where $m_D= v_2 Y_\nu/\sqrt{2}$, $m_R=\sqrt{2} v_s Y_R$ and $m_L=\sqrt{2} v_s Y_L$.
 For $m_L,m_R \ll m_D,\hat{M}_N$, the neutrino masses can be approximately written as
\begin{eqnarray}
m_{\nu_\ell} & \simeq & \frac{m^2_D\, m_L}{\hat{M}_N^2+m_D^2},\label{mnul}\\
m_{\nu_{H,H'}} & \simeq & \frac{1}{2}\left(\frac{\hat{M}_N^2\,m_L}{\hat{M}_N^2+m_D^2}+ m_R\right) \mp \sqrt{\hat{M}_N^2+m_D^2}\,.
\end{eqnarray}
 For simplicity, we choose $Y_{R}  = 0$. The ${\cal O}(0.01)$~eV neutrino mass can be achieved with the choice of parameters $Y_\nu \sim{\cal O}(0.1)$, $\hat{M}_N \sim 1$~TeV and $m_L \sim  {\cal O}(10^{-4})$~GeV \cite{Abdallah:2021npg}.

 The Yukawa terms in the Lagrangian that provide masses and the mixing for the quarks and leptons are written as follows 
\begin{eqnarray}
\lag \supset -Y_l^{ij} \bar l_{Li} H_1^\mathcal{C} e_{Rj} - Y_d^{ij} \bar q_{Li} H_1^\mathcal{C} d_{Rj} - Y_u^{ij} \bar l_{Li} H_1 u_{Rj}  + {\rm h.c.}.
\end{eqnarray}
To get the correct masses and the mixing for the charged quarks and leptons as in the SM their Yukawa couplings with $H_1$ need to be adjusted as $v_1 < v$. This can be achieved by considering $Y_f = \frac{Y_f^{\rm SM}}{\cos\beta}$, where $Y_f^{\rm SM}=\dfrac{\sqrt{2}m_f}{v}$. The expression for the coupling of the scalars with SM fermions is listed in Table.\ref{tab:hffCoupling}. 
\begin{table}[h!]
\begin{tabular}{|c|c|c|c|}
\hline
Couplings for   & ${h_i-f-\bar f}$ & $A-f-\bar f$ & $H^\pm-f-\bar{f'}$ \\
\hline
 $g_f$ & $Y_f^{\rm SM}\dfrac{Z_{i1}^h}{\cos\beta}$ & $Y_f^{\rm SM}\dfrac{Z_{i1}^A}{\cos\beta}$ & $Y_f^{\rm SM} \tan\beta$\\
\hline
\end{tabular}
\caption{The coupling of the fermions with different scalars of the model. $Z^h$ and $Z^A$ are the orthogonal matrices that diagonalize the CP-even and CP-odd scalar mass matrices.}
\label{tab:hffCoupling}
\end{table}

 The scalar potential of the model is given by 
 \begin{eqnarray}
     V(H_1,H_2,S,S_2) = &&  \mu_1 H_1^\dagger H_1 + \mu_2 H_2^\dagger H_2   
   + \mu_s S^\dagger S +\mu_{{s_{2}}}  {{S_{2}}}^\dagger {S_{2}} \nonumber\\
   &+&   \lambda_1 (H_1^\dagger H_1)^2 + \lambda_2 (H_2^\dagger H_2)^2  
             + \lambda_s (S^\dagger S)^2 + \lambda_{{s_{2}}} ({S_{2}}^\dagger {S_{2}})^2 \nonumber \\  
   &+&   \lambda_{12}( H_1^\dagger H_1) (H_2^\dagger H_2) + \lambda_{1s} (H_1^\dagger H_1) (S^\dagger S) + \lambda_{2s} (H_2^\dagger H_2) (S^\dagger S) \nonumber \\
  &+& \lambda_{1{s_{2}}} H_1^\dagger H_1 {S_{2}}^\dagger {S_{2}}  + \lambda_{2{s_{2}}} H_2^\dagger H_2  {S_{2}}^\dagger {S_{2}} + \lambda_{s{s_{2}}} S^\dagger S  {S_{2}}^\dagger {S_{2}}
	\nonumber \\
  &+& \lambda'_{12} \left|H_1^\dagger H_2\right|^2 - \{\mu_{1 2{s_{2}}} {S_{2}}^\dagger H_1^\dagger H_2 + \mu_{{s_{2}} s} S_{2}^2 S - \lambda_{T} {S_{2}} S H_1^\dagger H_2 + h.c\} 
	. 
 \end{eqnarray}
The inclusion of the scalar \(S_2\) in the model is essential. Without this field and specifically with its assigned $U(1)_X$ charge, a real pseudoscalar Goldstone boson would emerge due to the presence of a global $U(1)$ symmetry in the scalar potential where $\phi \to \phi\, e^{-i\theta Q}$ for $\phi \in \{H_1, H_2, S\}$.  When these scalars acquire their vacuum expectation value this symmetry is spontaneously broken resulting in a massless pseudoscalar in the model. The scalar $S_2$ breaks this global symmetry explicitly given by the coefficient of the $\mu_{12s_2}$ and $\lambda_T$ terms, preventing the emergence of the massless pseudoscalar. The vev's can be obtained by minimizing the potential, which gives rise to the following tadpole equations:

\begin{eqnarray}
	&&\mu_1  + \lambda_1 v_1^2 + \frac{\lambda_{12}+\lambda_{12}'}{2} v_2^2 + \frac{\lambda_{1s}v_s^2+\lambda_{1s_2}v_{s_2}^2}{2}  + \frac{v_2 v_{s_2}}{2v_1}\left(\lambda_T v_s-\sqrt{2}\mu_{12s_2}\right) = 0 \, , \\
	&&\mu_2 + \lambda_2 v_2^2 + \frac{\lambda_{12}+\lambda_{12}'}{2} v_1^2 + \frac{\lambda_{2s}v_s^2+\lambda_{2s_2}v_{s_2}^2}{2}  + \frac{v_1 v_{s_2}}{2v_2}\left(\lambda_T v_s-\sqrt{2}\mu_{12s_2}\right) = 0 \, , \\
	&&\mu_s + \lambda_s v_s^2 + \frac{\lambda_{1s}}{2} v_1^2 + \frac{\lambda_{2s}}{2} v_2^2 + \frac{\lambda_{ss_2}}{2} v_{s_2}^2+ \frac{v_1v_2 v_{s_2}}{2v_s}\lambda_T-\frac{v^2_{s_2}}{\sqrt{2}v_s}\mu_{s_2s} = 0 \, , \\
	&&\mu_{s_2} + \lambda_{s_2} v_{s_2}^2 + \frac{\lambda_{1s_2}}{2} v_1^2 + \frac{\lambda_{2s_2}}{2} v_2^2 + \frac{\lambda_{ss_2}}{2} v_{s}^2+  \frac{v_1 v_2}{2v_{s_2}}\left(\lambda_T v_s-\sqrt{2}\mu_{12s_2}\right)-\sqrt{2}\mu_{s_2s}v_s = 0 \, .\label{eqn:tpole}
\end{eqnarray}

The scalar fields can be expanded around the vev's, and written in the following form
\begin{eqnarray}
	\hspace*{-0.7cm}H_1 = \begin{pmatrix} \dfrac{v_1+\rho_1+i\eta_1}{\sqrt 2} \\ \phi_1^-\end{pmatrix}, \,  
	H_2 = \begin{pmatrix} \dfrac{v_2+\rho_2+i\eta_2}{\sqrt 2} \\ \phi_2^-\end{pmatrix}, \,  
	S   = \dfrac{v_s+\rho_s+i\eta_s}{\sqrt 2}, \,
	{S_{2}}   = \dfrac{v_{s_{2}}+\rho_{s_{2}}+i\eta_{s_{2}}}{\sqrt 2}.
\end{eqnarray}
After the spontaneous breaking of the electroweak and \(U(1)_X\) symmetry, the particle spectrum of the model contains four CP-even, two CP-odd, and a pair of charged scalars. Collecting the relevant quadratic terms and imposing the tadpole equations we can write down the mass matrix corresponding to the charged scalars in $(\phi_1^-\ \phi_2^-)^T$ basis as described in Eq.(\ref{Eq:Hcmassmatrix}) 
\begin{eqnarray}
M_\pm^2 = \left(\frac{\mu_{12}}{2v_1 v_2}-\frac{\lambda_{12}'}{2}\right) \begin{pmatrix}
v_2^2   & -v_1 v_2 \\
-v_1 v_2 & v_1^2   \\	\end{pmatrix} \, .
\label{Eq:Hcmassmatrix}
\end{eqnarray}
Here we define $\mu_{12}$ as $v_{s_2}( \sqrt{2} \mu_{12{s_{2}}} - v_s \lambda_{T}) $. The charged scalar mass matrix can be diagonalized by an orthogonal transformation with an angle $\beta = \arctan ({\frac{v_2}{v_1}})$. One of the charged scalars is a Goldstone boson which is eaten by the $W$ boson while the other one remains physical with mass given by 
\begin{eqnarray}\label{csmass}
M_{H^\mp}&=&\sqrt{\left(\frac{\mu_{12}}{2v_1 v_2}-\frac{\lambda_{12}'}{2}\right)}\,v,
\end{eqnarray}
where $v=\sqrt{v_1^2+v_2^2} \simeq 246$~GeV. The physical charged scalar is denoted as $H^{\mp} = - \sin{\beta} \phi^{\mp}_{1} + \cos{\beta} \phi^{\mp}_{2}$, which is orthogonal to the charged Goldstone boson. 

The mass matrix for the CP odd scalars in the basis $(\eta_1,\eta_2,\eta_s,\eta_{s_2})$ is given by
\begin{eqnarray}
	M_A^2 = \begin{pmatrix}
		\frac{v_2}{2v_{1}} \mu_{12} &~&- \frac{1}{2}\mu_{12} &~& \frac{1}{2} v_2 v_{s_2} \lambda_T &~& \frac{v_2}{2} \bar{\mu}_{12}\\
		- \frac{1}{2}\mu_{12}&~&\frac{v_1}{2v_{2}} \mu_{12} &~&- \frac{1}{2} v_1 v_{s_2} \lambda_T &~& -\frac{v_1}{2} \bar{\mu}_{12}  \\
		\frac{1}{2} v_2 v_{s_2} \lambda_T &~& - \frac{1}{2} v_1 v_{s_2} \lambda_T &~& \mu_{{s_{2}} s}\frac{v^2_{s_2}}{\sqrt{2}v_s} - \lambda_T \frac{v_1 v_2 v_{s_2}}{2 v_s} &~& -\frac{1}{2} v_1 v_{2} \lambda_T+\sqrt{2} \mu_{{s_{2}} s} v_{s_2}\\
		\frac{v_2}{2} \bar{\mu}_{12} &~&  -\frac{v_1}{2} \bar{\mu}_{12}  &~& -\frac{1}{2} v_1 v_{2} \lambda_T+\sqrt{2} \mu_{{s_{2}} s} v_{s_2}&~& \frac{\mu_{12{s_{2}}}v_1 v_{2}}{\sqrt{2}v_{s_2}}-\lambda_T \frac{v_1 v_2 v_{s}}{2 v_{s_2}}\!+\!2\sqrt{2}\mu_{{s_{2}} s} v_{s}
	\end{pmatrix} \label{eqn:Amass} ,\nonumber \\
\end{eqnarray}
where $\bar{\mu}_{12} = ( \sqrt{2} \mu_{12{s_{2}}} + v_s \lambda_{T})$. Diagonalizing the CP-odd scalar mass matrix, we find two zero eigenvalues corresponding to Goldstone bosons, which are absorbed by the \( Z \) and \( Z' \) bosons, rendering them massive. 

The CP-odd scalar mass eigenstates can be written as linear combinations of the interaction eigenstates as
\begin{eqnarray}
A_i = Z_{ij}^A \,\, \eta_j \,\, ,
\end{eqnarray}
where $Z_{ij}^A$ represents the mixing matrix for the CP-odd states.

Similarly, we find the following mass matrix for $CP$-even scalars using tadpole equations in the basis $(\rho_1,\rho_2,\rho_s,\rho_{s_{2}})$
\begin{eqnarray}
	M_H^2 = \begin{pmatrix}
		m_{11}^2 &~& m_{12}^2 &~& m_{13}^2 &~& m_{14}^2 \\
		m_{12}^2 &~& m_{22}^2 &~& m_{23}^2 &~& m_{24}^2   \\
		m_{31}^2 &~& m_{23}^2 &~& m_{33}^2 &~& m_{34}^2  \\
		m_{14}^2 &~&  m_{24}^2  &~& m_{34}^2 &~& m_{44}^2
	\end{pmatrix},\label{eqn:hmass} 
\end{eqnarray}
where
\begin{align*}
    &m_{11}^2 &=&\ 2\lambda_1 v_1^2 + \frac{v_2}{2v_{1}} \mu_{12} , &
    &m_{22}^2 &=&\ 2\lambda_2 v_2^2 + \frac{v_1}{2v_{2}} \mu_{12} ,& \\	
    & m_{33}^2 & =&\ 2\lambda_s v_s^2 + \frac{1}{\sqrt{2}}\mu_{{s_{2}} s} \frac{v_{{s_{2}}}^{2}}{v_s}-\frac{v_1 v_2v_{s_{2}} \lambda_{T}}{2 v_s},&
    & m_{44}^2  &=&\  2 \lambda_{s_{2}} v_{s_{2}}^2 + \mu_{12{s_{2}}}\frac{v_2 v_1}{\sqrt{2} v_{s_{2}}} -\frac{v_1 v_2 v_s \lambda_{T}}{2 v_{s_{2}}},& \\
    & m_{14}^2 & =&\ \lambda_{1{s_{2}}}\,v_1 v_{s_{2}} + v_2(\frac{1}{2} v_s \lambda_{T} - \frac{1}{\sqrt{2}} \mu_{12{s_{2}}}),&
    &m_{13}^2  &=&\ \lambda_{1s}\,v_1 v_s +\frac{1}{2} v_2 v_{s_{2}} \lambda_{T},& \\
    & m_{12}^2 & =&\ (\lambda_{12}+\lambda'_{12})v_1 v_2 -\frac{1}{2}v_{{s_{2}}} (\sqrt{2} \mu_{12{s_{2}}} - v_s \lambda_{T}),&
    &m_{23}^2 &=&\ \lambda_{2s}\,v_2 v_s+\frac{1}{2} v_1 v_{s_{2}} \lambda_{T},& \\
     &  & =&\ (\lambda_{12}+\lambda'_{12})v_1 v_2 + (M^2_A)_{12},&
    & &=&\ \lambda_{2s}\,v_2 v_s+ v_1 v_{s_{2}} \lambda_{T}+(M^2_A)_{23},& \\
    & m_{24}^2 & =&\ \lambda_{2{s_{2}}}\,v_2 v_{s_{2}} + v_1(\frac{1}{2} v_s \lambda_{T} - \frac{1}{\sqrt{2}} \mu_{12{s_{2}}}),&
    &m_{34}^2  &=&\ \frac{1}{2} v_1 v_2 \lambda_{T} + (\lambda_{s{s_{2}}} v_s - \sqrt{2} \mu_{{s_{2}} s}) v_{{s_{2}}}\,,& \\
     &  & =&\ \lambda_{2{s_{2}}}\,v_2 v_{s_{2}} + v_1 v_s \lambda_{T}+ (M^2_A)_{24}.&
    & & & & 
\end{align*}
Diagonalizing the mass matrix, the mass eigenstates can be written as linear combinations of the interaction eigenstates as
\begin{eqnarray}
h_i = Z_{ij}^h \,\, \rho_j \,\, ,
\end{eqnarray}
where $Z_{ij}^h$ represents the diagonalizing orthogonal matrix for the CP-even states. We denote these four mass eigenstates as $h_1$, $h_2$, $h_3$ and $h_4$. The mass eigenstate $h_{1}$ can be identified as the observed SM Higgs. This can be achieved by considering a small value for the $\tan\beta=\frac{v_2}{v_1} \leq \mathcal{O}(0.01)$ and scalar quartic couplings of $H_1$ with the other scalars. As a result, the $h_1$ scalar dominantly comes from the $H_1$ doublet. For the higher $\tan\beta$ value, to obtain the SM like Higgs we only keep the mixing of $\rho_1$ with $\rho_2$ and work in the alignment limit. The alignment limit condition for $h_1$ to be the SM is chosen by keeping the mixing angle between $\rho_1$ and $\rho_2$ equal to $\beta$. The mixing of  $\rho_1$ and  $\rho_2$ with  $\rho_3$ can be made small by choosing minimal values for $\lambda_{T}$, $\lambda_{1s}$ and $\lambda_{2s}$. Similarly the cross term between of $\rho_1$ and  $\rho_2$ with  $\rho_4$ can also be made negligible by choosing appropriate values of $\lambda_{1s2}$ and $\lambda_{2s2}$ to cancel the term containing $\mu_{12s_2}$ terms respectively. With this choice of parameters, the SM Higgs scalar mass and couplings are obtained accordingly.

\begin{figure}[h!]
\begin{center}
\includegraphics[width=0.49\textwidth]{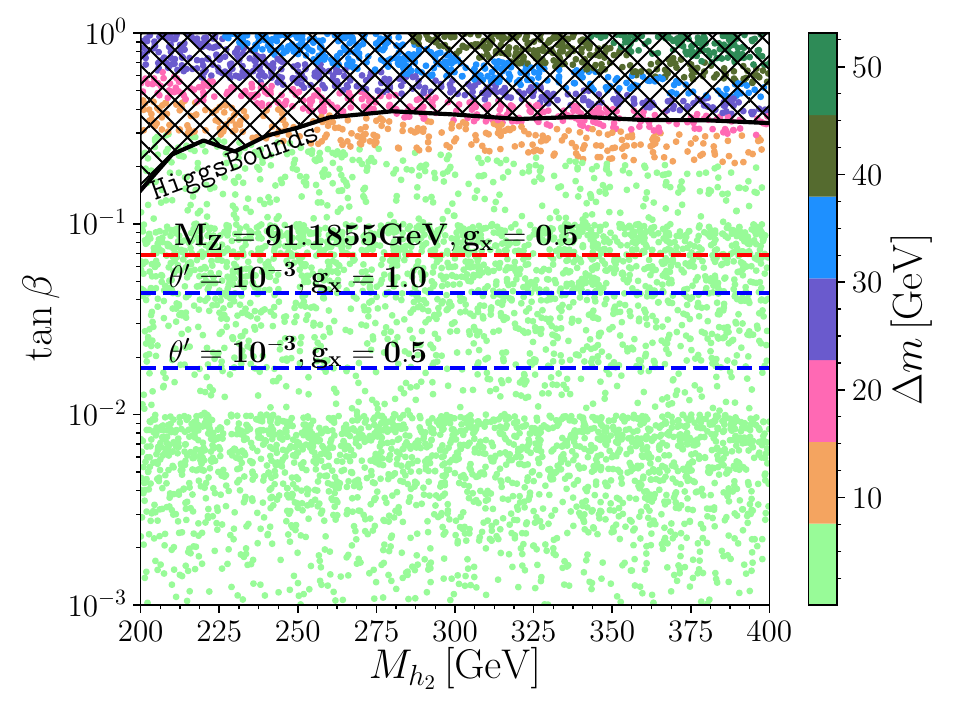}
\includegraphics[width=0.49\textwidth]{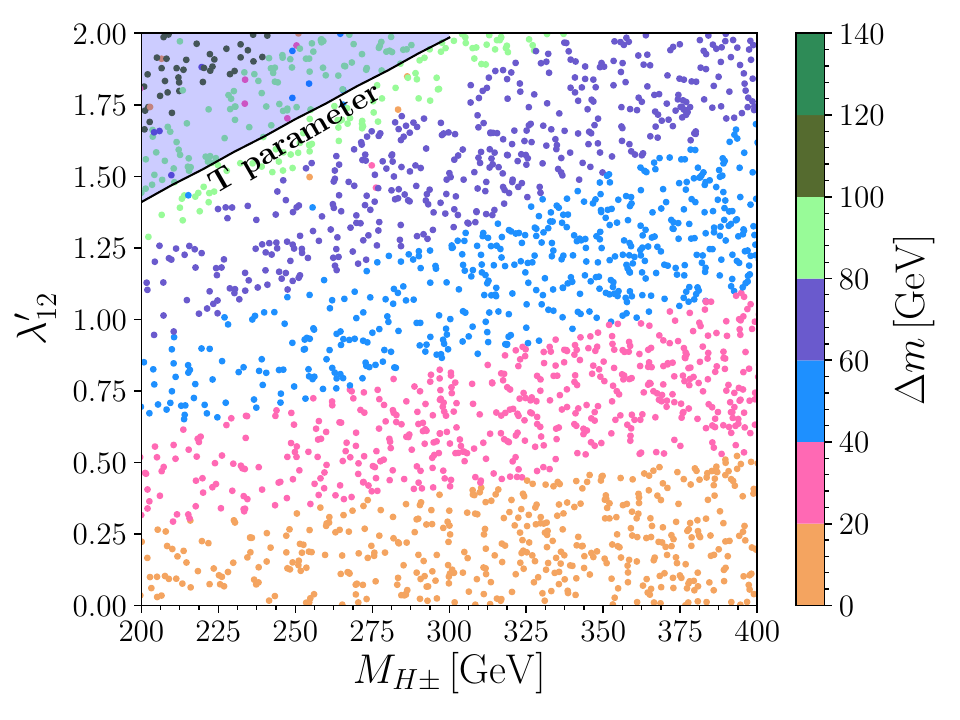}
\end{center}
\vspace{-0.5cm}
\caption{The mass difference between the pseudo scalar $A_2$ and the CP even scalar $h_2$ as a function of $\tan\beta$ and $M_{h_2}$ (left). The mass difference between the charged Higgs boson $H^\pm$ and the CP even Higgs boson $h_2$ as a function of $\lambda'_{12}$ and $M_{H^\pm}$ (right). }
\label{Fig-scalarmass}
\end{figure}

In the left plot of Fig.\ref{Fig-scalarmass}, we show the mass difference between the pseudoscalar $A_2$ and the CP-even scalar $h_2$. The plot is generated by making the mass mixing elements of $\rho_s$, $\rho_{s_2}$ with $\rho_1$, $\rho_{2}$ to be vanishing small by choosing an appropriate relation between the parameters as discussed above. 
In this scan, we have varied $\tan \beta \in [10^{-3},1.0]$ and $M_{h_2} \in [200,400]$ GeV and the parameters $\lambda_1$, $\lambda_2$ and $\mu_{12s_2}$ are determined by the $M_{h_1}$, $M_{h_2}$ and mixing angle $\beta$.  In addition, we have fixed $v_{s} = 10$ GeV and $v_{s_2} = 229$ GeV, respectively. The motivation for choosing such a hierarchical vev for the singlet scalars is to ensure that all the singlet scalars are heavier than the doublet scalars. This can be understood from the explicit form of $m^{2}_{33}$ and  $m^{2}_{44}$.  From the plot, one can see that for larger values of $\tan\beta$  a significant mass difference between the $A_2$ and $h_2$ can be possible. The mass difference is controlled primarily by the $\lambda_2$ parameter. Here we have fixed $\lambda_{12}'$ to be small so that the $H^\pm$ mass will be nearly equal to the $h_{2}$ mass. For this plot, the SM Higgs properties are ensured by working in the alignment limit. In the general 2HDM, when working in the alignment limit we find that the cubic coupling between the second CP even Higgs and the two gauge bosons $(h_2 \, W^+W^-, h_2 \, ZZ )$ does not exist. However, in the present scenario, as the $H_2$ is charged under the $U(1)_X$ gauge symmetry, $h_2ZZ$ coupling can be induced due to the presence of $Z-Z'$ mixing. As the $\theta'$ increases with an increase in $\tan\beta$, for a large value of $\tan\beta$ the $h_2ZZ$ coupling increases. Therefore existing searches on BSM scalars restrict our parameter space, which we have shown by the solid black line in the plot. We find that search channels such as $ p p \to h_2 \to 4\ell,2\ell 2q, 2\ell \mET $ \cite{CMS:2018amk} and $ gg \to A_2 \to Z h_1 \to 2\ell 2b $ \cite{CMS:2019qcx} constrains our parameter these search channels are incorporated in the public package {\tt HiggBounds} \cite{Bechtle:2011sb}. Thus the hatched region is disallowed by the existing searches on the BSM scalars. The large mixing angle $\theta'$ can also alter the $h_1ZZ$ coupling, which can get constraints from the Higgs signal. We find that this constraint is less stringent than the {\tt HiggBounds} constraint.  
Due to mixing between $Z$ and $Z'$, the mass of the $Z$ boson deviates from its precisely measured SM value, $M_Z = 91.1876 \pm 0.0021$ GeV \cite{ParticleDataGroup:2020ssz}. The red dashed line represents the constraint on $\tan\beta$ imposed by the measurement of the mass of the $Z$ boson. As the mixing angle $\theta'$ increases with an increase in $\tan\beta$, the region above the red dashed line is excluded because the $Z$ boson mass drops below its observed lower limit. The most stringent limit on the parameter space comes from the $Z-Z'$ mixing angle $\theta'$. In the plot, the blue dashed lines imply $\theta' = 10^{-3}$ for two values of $g_{x} = 0.5$ and $1.0$. Therefore, the $\theta '$ constraint rules out the region above the blue lines. All the scanned points are with $g_{x} = 0.5$, for $g_{x} = 1.0$ we showed the value of $\tan \beta$ at which $\theta'$ reaches $10^{-3}$ radian. Therefore, to satisfy the $\theta'$ constraint $\tan\beta \lesssim 10^{-2}$ is required. 

For $\tan{\beta} << 1$, the diagonal element in the second row of the mass matrices of the CP even and CP odd scalars is much larger than the off-diagonal entries. Therefore, the second eigenvalue of these mass matrices will be nearly equal to the $(2,2)$ elements of these matrices. Comparing the second row for the mass matrices of CP even and CP odd scalar, we find that the $(2,2)$ elements of these matrices are nearly equal for $\tan{\beta} << 1$, and therefore $h_2$ and $A_2$ will be nearly degenerate in mass. In the limit $\lambda_T \to 0$, $v_2 <<v$ and $v_s << v_{s_2}$ the masses of the two physical CP odd scalars become  
\begin{eqnarray}\label{asmass}
M_{A_{2}}^2&\simeq&\frac{\mu_{12s_2}v_{s_2} v^2}{\sqrt{2}v_1 v_2} \left(1+\frac{v^2_1 v^2_2}{v^2_{s_2} v^2}\right) \nonumber \\
M_{A_{s}}^2&\simeq& \frac{\mu_{s_2s}}{\sqrt{2} v_s} \left(4 v_s^2+v_{s_2}^2\right).
\end{eqnarray} 
The scalar $A_{2}$  predominantly comes from the CP odd part of the second Higgs doublet while the $A_{s}$ is mainly comprised of the CP odd part of the singlet scalars. In this limit, $h_2$ is predominantly from the $H_2$ doublet with mass $m_{h_2} \simeq \sqrt{\mu_{12} \cot{\beta}}$, while the other two mass eigenstates, $h_3$ and $h_4$ are linear combinations of the two singlet scalars $S$ and $S_2$.

In the right plot, we show the mass difference between the charged Higgs boson $H^\pm$ and the CP even Higgs boson $h_2$ as a function of $\lambda'_{12}$ and $M_{H^\pm}$. Note that $\lambda_{12} \geq -2 \sqrt{\lambda_1 \lambda_2 } $ and $\lambda'_{12} + \lambda_{12} \geq -2 \sqrt{\lambda_1 \lambda_2} $ to satisfy the stability of the scalar potential \cite{Barik:2024mkr}. By choosing a suitable value of $\lambda'_{12}$ the charged Higgs can be made heavier or lighter than $h_2$ and $A_2$. In this scan we have consider $\tan\beta = 10^{-2}$ with $v_{s} = 10$ GeV and $v_{s_2} = 229$ GeV. For the given value of $\tan\beta$, the mass difference between $h_2$ and $A_2$ as seen from the left plot is small. The scanned points with $\tan\beta = 10^{-2}$ are allowed by the {\tt HiggBounds}, {\tt HiggSignals} \cite{Bechtle:2011sb,Bechtle:2013xfa} and $Z-Z'$ mixing angle constraints. We find that the electroweak precision observable T-parameter restricts our parameter space when the mass difference between $M_{H^\pm}$ with $M_{h_2}$ and $M_{A_2}$ is nearly $\approx 90$ GeV. The measured value of the T-parameter is $T = 0.09\pm 0.14$ \cite{Haller:2018nnx} and the T-parameter constraint rules out the blue-shaded region.
 

\section{BSM Scalar searches}
 \label{sec:scalar search}
In this work, we study the signals from the BSM scalars $h_2, A_2$ and $H^\pm$  that couple very weakly with the SM fermions when the $Z'$ is light. As the coupling strength depends on $\tan{\beta}$, the upper bound of $10^{-2}$ does not allow any SM fermion to have large couplings with the 2HDM scalars. Hence their direct production in association with the heavy quark flavors ($t\bar{t}h_2, tbH^-$) is negligibly small, so the constraints from the existing scalar searches at LHC will not be applicable in this scenario. However, these scalars can be produced in pairs at LHC, mediated by the SM gauge bosons. When these scalars are produced at LHC they can decay into SM particles and more interestingly into BSM modes too. This work focuses on the scalars decaying into the mode containing the $Z'$. The relevant couplings of these scalars with gauge bosons have been listed in Table.\ref{tab:feynrules}. 
\begin{table}[t!]
    \centering
    \begin{tabular}{|c|c|}
    \hline
       Vertices  & Couplings  \\
       \hline 
        $H^{\pm}W^{\mp}_{\mu}Z'_{\nu}$  & $-ig_2g_{x}v_{2}\cos\beta g_{\mu\nu}$   \\
       \hline
        $h_{2}Z'_{\mu}Z'_{\nu}$ & $2i\left[(4v_{s}Z^{h}_{23} + v_{s_{2}}Z^{h}_{24})g^{2}_{x} + v_{2}g^{2}_{x}Z^{h_{2}}_{22}\right]g_{\mu\nu}$ \\
        \hline
        $h_{2}Z_{\mu}Z'_{\nu}$ & $-iv_{2}Z^{h}_{22}\left[g_{x}g_{1}\sin\theta_{W} + g_{2}g_{x}\cos\theta_{W} \right]g_{\mu\nu}$ \\
        \hline
        $H^{\pm}A_{2}W^{\mp}_{\mu}$ & $\frac{g_{2}}{2}\left[\cos\beta Z^{A}_{22} - \sin\beta Z^{A}_{21}\right]\left(p_{A_{2}} - p_{H^{\pm}}\right)_{\mu}$ \\
        \hline
        $H^{\pm}h_{2}W^{\mp}_{\mu}$ & $\frac{i g_{2}}{2}\left[\cos\beta Z^{h}_{22}-\sin\beta Z^{h}_{21}\right]\left(p_{h_{2}} - p_{H^{\pm}}\right)_{\mu}$ \\
        \hline
        $A_{2}h_{1}Z'_{\mu}$ & $\left[g_{x}\left(2Z^{A}_{23}Z^{h}_{13} - Z^{A}_{24}Z^{h}_{14}\right) - g_{x}Z^{A}_{22}Z^{h}_{12}\right]\left(p_{A_2} - p_{h_{1}}\right)_{\mu}$ \\
        \hline 
        $A_{2}h_{2}Z_{\mu}$ & $\frac{1}{2}\left(g_{1}\sin\theta_{W} + g_{2}\cos\theta_{W}\right)\left[Z^{A}_{21}Z^{h}_{21} + Z^{A}_{22}Z^{h}_{22}\right]\left(p_{A_2} - p_{h_{1}}\right)_{\mu}$ \\
        \hline
    \end{tabular}
    \caption{The coupling of the scalars $H^{\pm}$, $h_2$ and $A_2$ with gauge bosons.}
    \label{tab:feynrules}
\end{table}

\begin{figure}[t!]
\begin{center}
\includegraphics[width=0.495\textwidth]{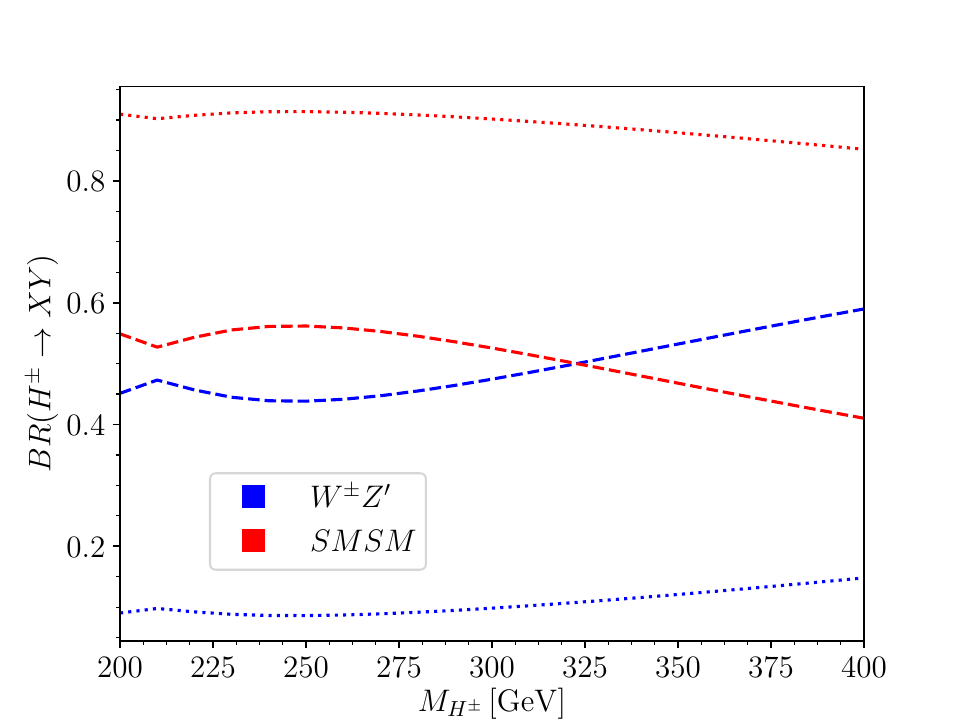}
\includegraphics[width=0.495\textwidth]{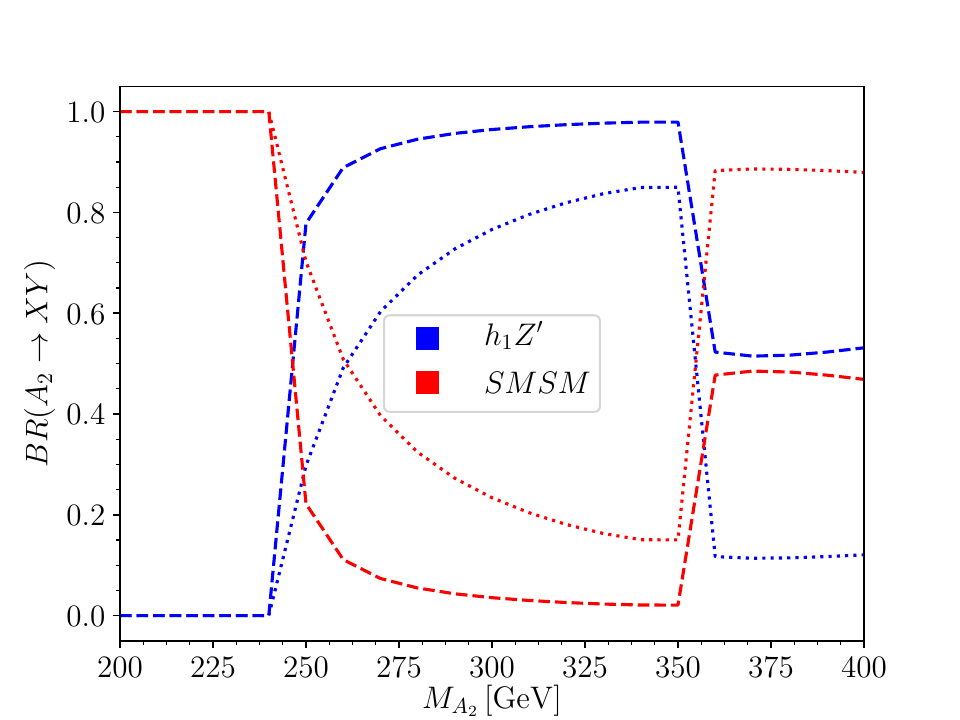}\\
\includegraphics[width=0.495\textwidth]{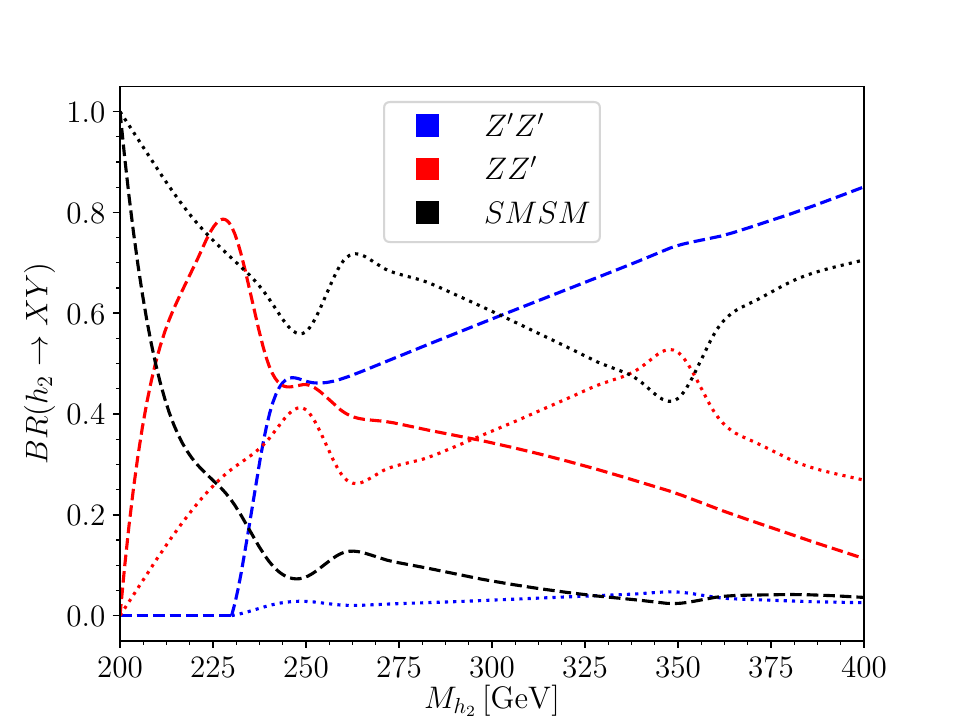}
\end{center}
\vspace{-0.5cm}
\caption{Branching ratio of BSM scalars to different decay modes. The dashed and dotted lines correspond to $g_x = 0.575$ and $g_x = 0.2$ respectively with $M_{Z'} = 115$ GeV. }
\label{H2decay}
\end{figure}
Now we focus on various decay channels of the BSM scalars $H^\pm$, $A_2$ and $h_2$. The expressions for the decay widths of these scalars to various channels are given in the Appendix.\ref{app:decaywidth}. In Fig.\ref{H2decay} we show the branching ratio (BR) of BSM scalars to SM modes and the modes containing at least one $Z'$ for two values of the gauge coupling $g_x = 0.575$ and $g_x = 0.2$ with $M_{Z'} = 115$ GeV. We assume the BSM neutrinos to be heavier than the scalars $H^\pm$, $A_2$, and $h_2$ and are therefore not kinematically accessible as decay modes. The left plot in the top panel of the figure shows that $H^{+}$ decays to the $W^+ Z'$ with around 10\% probability for $g_x = 0.2$.  Its BR to this mode increases with $g_{x}$, as the interaction strength is proportional to $g_x$. The mixing with the SM scalar drives the SM decay modes and it mostly decays to $t \bar{b}$ in the SM modes. The small dip in the BR of $H^+ \to W^+ Z'$ at $M_{H^\pm} \approx 205$ GeV is due to the opening of the $h_1 W^+$ decay mode. Note that the BSM mode will increase for lighter $Z'$ mass for a fixed value of $g_x$. From the top right plot of the figure, we find that the pseudoscalar $A_{2}$ completely decays to the SM mode which is represented by the sum of $b \bar{b}$, $g g$ and $h_{1} Z$ modes when the $h_1 Z'$ is kinematically disallowed. As the $h_1 Z'$ decay mode becomes accessible to the $A_2$, this mode takes over and remains dominant until the $t\bar{t}$ channel opens up at $M_{A_2} > 350$ GeV. In the bottom plot of the figure, we have shown the corresponding decay channels of the CP even scalar $h_2$. The $h_2$ decay to SM modes comes mostly from its decay to $W^+ W^-$ and $Z Z$ when the $ZZ'$ and the $Z'Z'$ modes are not kinematically open. As these modes open up, they begin dominating the decay probabilities of $h_2$, and the SM decay modes are suppressed. The slight rise seen in the BR of the SM mode at $M_{h_2} \approx 250$ GeV and $M_{h_2} \approx 350$ GeV is due to the opening of additional decay modes ($h_1 h_1$ and $t\bar{t}$) which becomes kinematically allowed. Note that the decay mode of these BSM scalars containing at least a $Z'$ increase with $g_x$ as these scalars carry $U(1)_X$ charge.  

\begin{table}[h!]
\begin{center}\scalebox{1.00}{
\begin{tabular}{|c|c|c|c|c|}
\hline   $\rm{BR}^{Z'} _{d \bar{d}}~$ &$\rm{BR}^{Z'} _{u \bar{u}}~$  &$\rm{BR}^{Z'} _{\nu_{\ell} \nu_{\ell}'}$~&$\rm{BR}^{Z'} _{{\ell^+} \ell^-}$~&$\rm{BR}^{Z'} _{{\tau^+} \tau^-}$\\ \hline
     0.350 & 0.232&0.210&0.138&0.069        \\ 
\hline
 \end{tabular}}
\end{center}
\caption{BR of the $Z'$ to different decay modes for two values with $g'_{x} = 10^{-4}$, $\tan\beta = 10^{-4}$ and $M_{Z'} = 115$ GeV. The $\ell$ corresponds to the electron and muon.}
\label{tab:Zp_decay}
\end{table}

The processes shown in Table.\ref{Tab:sigprod} are the relevant channels for producing the BSM scalars at the LHC. After all the decay cascades of the unstable particles culminate in stable states, we will be left with jets, leptons, and photons in the final state. The coupling of the $Z'$ with SM fermions and the formulas for its decay branching fraction to SM fermion modes are in the Appendix. (\ref{app:feynrule}) and (\ref{app:decaywidth}) respectively. The decay branching fraction of the new gauge boson $Z'$ to SM modes are shown in Table.\ref{tab:Zp_decay} for $M_{Z'} = 115$ GeV with $g'_{x} = 10^{-4}$ and $\tan\beta = 10^{-4}$. As at LHC, many searches for the BSM signal have been performed with various final states containing the stable SM particles, therefore it is necessary to check whether our parameter space is constrained from these searches.   

\begin{figure}[h!]
\begin{center}
\includegraphics[width=0.49\textwidth]{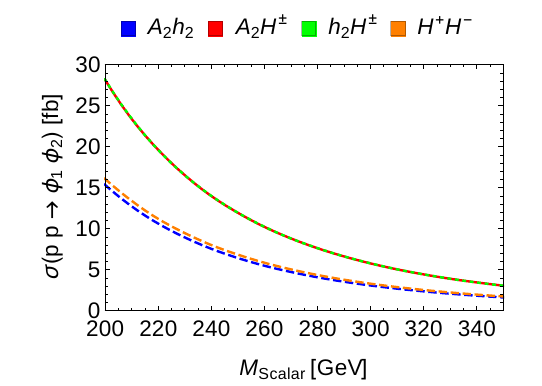}
\end{center}
\vspace{-0.5cm}
\caption{The cross-section of the processes $p p \to \phi_1 \phi_2$ at $\sqrt{s} = 13$ TeV, where $\phi_1, \phi_2 \in \{A_2, h_2, H^\pm\}$. The scalar masses $(A_2, h_2, H^\pm)$ are considered in the range $M_{\rm{Scalar}} \in [200, 350] \, \text{GeV}$.
}
\label{Fig-xsecH2H2}
\end{figure}

We find that the most stringent limits on our parameter space come from the $4\ell$ searches carried by the ATLAS collaboration \cite{ATLAS:2021kog} at the LHC with 13 TeV center of mass energy. This search measures the differential distribution of the SM $4\ell$ background for various observables with additional particles allowed to be present in the event. The selection criteria for the leptons $p_{T_{e(\mu)}} > 7(5)$ GeV, $p_{T_{\ell_1}} > 20$ GeV,  $p_{T_{\ell_2}} > 10$ GeV, $M_{\ell^- \ell^-} > 5$ GeV, $|\eta| < 2.4$, $\Delta R_{\ell\ell} > 0.05$, $\Delta R_{\ell\gamma} > 0.1$ and $\Delta R_{j\ell} > 0.3$. This analysis has been included in the public package {\tt Rivet-3.1.4} \cite{Buckley:2010ar}, which we have used to compare our results with the experimental findings. This package provides {\tt YODA} \cite{Buckley:2023xqh} files which are then taken as input for the package {\tt Contur} \cite{Buckley:2021neu} to yield constraints on our parameter space. To extract the bounds on our parameter space we evaluate the cross-section of the processes shown in Table.\ref{Tab:sigprod} at $\sqrt{s} = 13$ TeV at the LHC, and is plotted in Fig. \ref{Fig-xsecH2H2}.

\begin{figure}[h!]
\begin{center}
\includegraphics[width=0.49\textwidth]{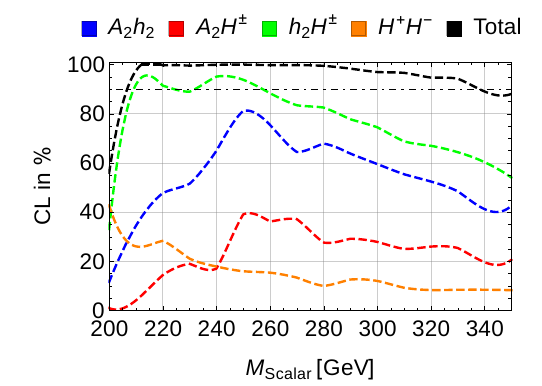}
\includegraphics[width=0.49\textwidth]{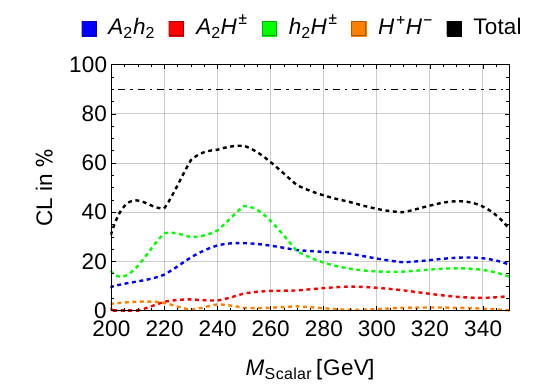}
\end{center}
\vspace{-0.5cm}
\caption{Exclusion plot from the ATLAS $4\ell + X$ searches for $M_{Z'} = 115 \, \text{GeV}$ with $g_{x} = 0.575$ (left plot) and $g_x = 0.2$ (right plot). The scalar masses $(A_2, h_2, H^\pm)$ are considered in the range $M_{\rm{Scalar}} \in [200, 350] \, \text{GeV}$. The black dot-dashed line corresponds to the exclusion limit at 90\% C.L.
}
\label{Fig-contur_cl}
\end{figure}

We find that the invariant mass of the opposite-sign-same-flavor (OSSF) leptons constrain our parameter space most. The exclusion limits on the parameter space obtained from the package {\tt Contur} have been plotted in Fig.\ref{Fig-contur_cl} for two values of the gauge coupling $g_x = 0.575$ (left plot) and $0.2$ (right plot) with $M_{Z'} = 115$ GeV.  From the left plot of the figure, we find that when the BR of $A_2$ and $h_2$ decay via the $Z'$ mode are small, the exclusion from the ATLAS $4\ell +X $ search is very weak. When all the scalar production modes containing $Z'$ decays open up, a much more stringent bound appears at $M_{\rm{Scalar}} \approx 240$ GeV. We find that for heavier scalar mass, the cross-section for all the relevant processes starts falling and therefore the bound from the $4\ell +X$ search also becomes weaker. The subprocess that contributes the most to the exclusion plot is the $p p \to h_2 H^{\pm}$, although it has an almost equal cross-section as the $p p \to A_2 H^{\pm}$ process. 
This happens because one can get a maximum of three $Z'$ bosons from $h_2 H^{\pm}$ production and the subsequent decay of the scalars. Therefore, it has the largest effective $4\ell + X $ cross-section. Similarly, the process $p p \to A_2 h_2$ that occurs via the $Z$-boson mediation is found to have the smallest cross-section among the four processes listed in Table.\ref{Tab:sigprod}. But the pseudoscalar $A_2$ decays to $h_1 Z'$ with very large BR (Fig. \ref{H2decay}) and therefore it gives the second largest contribution to the $4\ell+X$ final state as can be seen from Fig. \ref{Fig-contur_cl}. In the remaining two processes, the $p p \to H^{-} H^{+}$ has a lower production cross section and contributes least to the four lepton signals. The constraints can be evaded by varying the new gauge coupling $g_x$ as it controls the decay BR of the scalars into modes containing $Z'$. This effect can be seen from the right plot of Fig. \ref{Fig-contur_cl} with $g_x = 0.2$. For $Z'$ lighter than the $Z$ boson, the BR of the scalar decay modes containing $Z'$ is larger than that of the heavier $Z'$ case for a given value $g_x$ due to the phase space enhancement. Therefore the ATLAS $4\ell+X$ search can help put stringent constraints on light $Z'$ scenarios that appear as decay products of BSM scalars produced at the LHC. Similarly, this signal also forms a useful channel for looking at a scenario such as ours in the multilepton final states. 
We have also used the public package {\tt CheckMATE} \cite{Drees:2013wra} to find constraints on our model. We considered all the search channels encoded in this package and we find that only the multilepton channels with missing transverse energy, derived in the context of electroweakino (charginos and neutralinos) searches in supersymmetry appear to constrain our parameter space \cite{CMS:2017moi} mildly. However, these constraints are much weaker than those in Fig. \ref{Fig-contur_cl}.

We briefly comment on other possibilities for the scalar search in the model at the LHC. In the scenario where the heavy neutrinos are lighter than $A_2$, $h_2$ and $H^\pm$, these scalars can decay to the heavy neutrinos and the SM lepton via the $Y_{\nu}$ coupling. These decay modes are dominant when $Y_{\nu}$ is of the same order as the couplings listed in Table.\ref{tab:feynrules}, which govern the decay of the scalar to the mode containing $Z'$. Note that in this model the mixing angle between the light and heavy neutrinos $|V_{\ell N}|^2 \sim \left(\frac{m_{D}}{\hat{M}_{N}}\right)^2$ must be $< 10^{-5}$ for $\mathcal{O}(100)$ GeV heavy neutrino mass \cite{CMS:2018jxx}. 
For $\mathcal{O}(100)$ GeV heavy neutrino mass we find that $Y_\nu \sim \mathcal{O}(0.1)$ is allowed from this constraint with $v_2 \simeq 2$ GeV. The correct light neutrino masses for these points are obtained by choosing appropriate $Y_{L}$. The scalars of mass $\geq 200$ GeV with $Y_\nu \sim \mathcal{O}(0.1)$ decay to a heavy neutrino of mass $150$ GeV and a SM lepton with $\sim 0.98$, thus rendering the decay of these scalars to the modes containing $Z'$ to negligibly smaller BR of $\sim 1\%$ with $g_x = 0.575$. The heavy neutrinos decay to SM leptons associated with $W$, $Z$, $Z'$, and $h_{1}$. The branching fraction of heavy neutrinos to their various decay modes is shown in Table.\ref{tab:vH_decay} for two values of $g_x=0.575$ and $g_x=0.2$. The decay width of the process ${\nu_{H}} \to {\nu_{\ell} Z'}$ is directly proportional to $g^2_x$, therefore BR of heavy neutrinos to this mode increases with increase in $g_x$. When the scalars $A_2$, $h_2$, and $H^\pm$ decay to SM leptons and heavy neutrinos followed by the decay of heavy neutrinos to the stated modes can give rise to multilepton final states. The multilepton final state in the context of a neutrinophilic Higgs doublet model has been studied before by one of the authors \cite{Huitu:2017vye}, and so we do not consider analyzing this interesting channel in the present work. However, it is worth pointing out that in this model, we do find that when these scalars decay dominantly to the heavy neutrinos and SM lepton, it is severely constrained by the ATLAS multilepton search where the invariant mass of the four leptons constraints the parameter space with the lighter heavy neutrino states more stringently.    
\begin{table}[h!]
\begin{center}\scalebox{1.00}{
\begin{tabular}{|c|c|c|c|c|}
\hline  $g_x$ &  $\rm{BR}^{\nu_{H}} _{\ell^{\pm}W^{\mp}}~$ &$\rm{BR}^{\nu_{H}} _{\nu_{\ell} Z}~$  &$\rm{BR}^{\nu_{H}} _{\nu_{\ell} Z'}$~&$\rm{BR}^{\nu_{H}} _{\nu_{\ell} h_1}$\\ \hline
     0.575  & 0.535 & 0.230 &0.181&0.054        \\
     0.2  & 0.636 & 0.274 & 0.026&0.064     \\ 
\hline
 \end{tabular}}
\end{center}
\caption{BR of the heavy neutrinos to different decay modes for two values of $g_x$. The decay width of process ${\nu_{H}} \to {\nu_{\ell} Z'}$ increases with increase in gauge coupling $g_x$.}
\label{tab:vH_decay}
\end{table}

Another interesting feature of this model is that the $Z'$ can decay to SM leptons at one loop level with the heavy neutrinos and $A_2$, $h_2$ and $H^\pm$ running in the loop \cite{Abdallah:2021dul}. This mode becomes relevant when the GKM parameter is vanishingly small $(\sim 10^{-5})$ and the contribution solely comes from the one-loop decay, as this can exceed the tree-level total decay width. Therefore, the $Z'$ decays only to the charged leptons and light neutrinos and becomes completely leptophilic. One loop decay width of $Z'$ for this process depends on the Yukawa coupling $Y_{\nu}$, $g_x$, and the masses of the particles running in the loop. We find that when the loop decay is dominant, the $Z'$ can decay to charged leptons and the neutral lepton with equal probability. Each heavy neutrino couples with different flavors of the SM leptons, therefore, $Z'$ can decay to flavor violating modes e.g. $e^{\pm} \mu^{\mp}$, $e^{\pm} \tau^{\mp}$, etc. This flavor violating decay mode of $Z'$ can be up to $25$ \% which can be obtained by suitably choosing $Y_{\nu}$. Due to this large BR of $Z'$ to charged lepton the constraint from the above-mentioned multilepton searches becomes more stringent. However, with such a large BR to charged lepton modes heavier BSM scalar can be probed. Additionally, the invariant mass peak in the opposite-sign-different-flavor (OSDF) pair of leptons due to the presence of the $Z'$ will provide an extra variable to distinguish the signal from the SM background. The current mass range that we consider for the scalars and $Z'$ in this work is found to be ruled out by the ATLAS multilepton analysis unless the $g_x$ is chosen to be small.

\section{Collider Analysis}
\label{sec: collider}
We now evaluate the search prospect of the BSM scalars at the LHC arising from $H_{2}$ doublet. In the limit $\tan\beta \simeq 10^{-4}$, the dominant production mode of these scalars is presented in Table.\ref{Tab:sigprod}.  
After production, these BSM scalars can further decay into gauge bosons. To calculate the production cross-section for the signal at the 14 TeV run of LHC, we have considered all four production modes that are mentioned in Table.\ref{Tab:sigprod}. Note that the enhancement in the production rates of the different modes ranges from 15-20\% compared to the rates at $\sqrt{s}=13$ TeV of the LHC shown in Fig.~\ref{Fig-xsecH2H2}. For all BP's the CP-even scalar $h_2$ dominantly decays either via $ZZ'$ or $Z'Z'$ modes and the CP-odd scalar $A_{2}$ decays into $Z'h_{1}$ mode. In contrast to that, the $H^{\pm}$ branching ratio to $WZ'$ mode can vary depending upon the BP choice. 
In Table.\ref{tab:bps}, we show the branching ratio of these decay channels for different benchmark scenarios.
The gauge bosons that arise from the heavy scalar decays can give rise to different multilepton final states. For the present study, we only consider the more sensitive $4\ell+X$ final state and investigate the discovery prospect of this channel in terms of the BSM scalar searches. 
\begin{table}[t!]
\begin{center}
    \begin{tabular}{|c|c|c|}
    \hline
        Process & Mediator & Decay \\
        \hline
        $p p \to H^{+}H^{-}$ & $\gamma/Z$ & $H^{\pm} \to W^{\pm} Z'$ \\
        $p p \to H^{\pm}h_{2}$ & $W^{\pm}$ & $H^{\pm} \to W^{\pm} Z'$, $h_{2} \to Z'Z'/ZZ'/ZZ/WW$ \\
        $p p \to H^{\pm} A_{2}$ & $W^{\pm}$ & $H^{\pm} \to W^{\pm} Z'$, $A_{2} \to Z'h_{1}/Zh_{1}$\\
        $p p \to h_{2} A_{2} $ & $Z$ & $h_{2} \to Z'Z'/ZZ'/ZZ/WW$, $A_{2} \to Z'h_{1}/Zh_{1}$ \\
        \hline
    \end{tabular}
\end{center}
\caption{The production and decay modes for different BSM scalars that primarily belong to the $H_{2}$ doublet field, at the LHC. For their subsequent decay, we consider those modes that can give rise to $4\ell + X$ final states with appreciable branching ratio.}
\label{Tab:sigprod}
\end{table}
  
\begin{table}[h!]
\begin{center}\scalebox{1.00}{
\begin{tabular}{|c|c|c|c|c|c|c|c|c|c|c|c|}
\hline &  $\sigma_{\sum \phi_1 \phi_2}$ (fb) &  $M_{\phi_{1,2}}$ (GeV) & $g_x$ &  $\rm{BR}^{H^\pm} _{W^{\pm} Z'}~$ &$\rm{BR}^{h_2} _{Z' Z'}~$  &$\rm{BR}^{h_2} _{Z Z'}$~&$\rm{BR}^{h_2} _{Z Z}$~& $\rm{BR}^{h_2} _{W^+ W^-}~$& $\rm{BR}^{A_2} _{h_{1}Z'}~$& $\rm{BR}^{A_2} _{h_{1}Z}~$ & $C.L.(\%)$\\ \hline
     {\tt BP1} & 20.67      & 300         &0.3  & 0.197 & 0.109 &0.513&0.034&0.076&0.880&0.100&72.4         \\
     {\tt BP2} & 10.98      & 350         &0.4  & 0.354 & 0.318 & 0.564&0.010&0.021&0.957& 0.027& 68.2    \\ 
      {\tt BP3}   &   6.23        & 400    & 0.5     &0.520 & 0.609 & 0.274&0.002&0.003&0.461&0.005& 51.9   \\   
\hline
 \end{tabular}}
\end{center}
\caption{Masses, total cross-section of $p p \to \sum \phi_1 \phi_2$, where $\phi_1, \phi_2\in\{A_,h_,H^{\pm}\}$, BR of the scalars to relevant channels for three {\tt BP}s and C.L. exclusion limit from the ATLAS $4 \ell + X$ search. The mass of the $Z'$ is fixed to $M_{Z'}=115$ GeV for all BP's.}
\label{tab:bps}
\end{table}
\begin{table}[h]
\centering
\begin{tabular}{|c|c|c|c|c|c|c|c|c|c|c|c|c|}\hline
 ${\lambda_{1}}$ & $\lambda_{2}$ & $\lambda_{3}$ & $\lambda_{4}$ & $\lambda_{1s}$ & $\lambda_{2s}$  & $\lambda_{s}$&$\lambda_{s_2}$ &$\lambda_{ss_2}$&$\mu_{ss_2}$ (GeV)&$v_{s}$ (GeV) & $\tan\beta$ & $g'_{x}$  \\
\hline
 0.129 & 1.0 & 0.005 & 0.005 & $10^{-4}$ & $10^{-4}$  & $1.0$ & 3.0&1.0&150.0&10.0& $10^{-4}$& $10^{-4}$  \\ \hline 
\end{tabular}
\caption{Other relevant parameters of the model that we have fixed throughout the analysis.}
\label{tab:scalar}
\end{table}

To generate the signal events we have implemented the BSM model in the {\tt SARAH} \cite{Staub:2015kfa} package to obtain the {\tt Universal FeynRules File (UFO)} \cite{Degrande:2011ua}. The particle masses and decay width are generated using the package {\tt Spheno} \cite{Porod:2003um}. The dominating SM background corresponding to this final state comes from the following sub processes\footnote{The background process $p~p \to 4\ell+jets$ consists with three sub-processes $p~p \to 4\ell$, $p~p \to 4\ell + 1j$ and $p~p \to 4\ell + 2j$ where $p~p \to 4\ell$ is 
the most dominant background.} 
\begin{align*}
p p \to 4\ell+jets, &&  p p \to VVV, &&  p p \to t\bar{t}V/t\bar{t}h, &&   p p \to WZ/Vh \,\, ,
\end{align*}
where $V=W^\mp, Z$.
Both the signal and background events are generated using the {\tt Madgraph} (v2.6.7)  \cite{Alwall:2011uj} at the parton level while fixing the center of mass energy to 14 TeV. These parton level events were then passed to {\tt Pythia8} \cite{Sjostrand:2014zea} for the showering and hadronization. The detector effects were incorporated using the {\tt Delphes-3} \cite{deFavereau:2013fsa} simulator. Finally, the signal and background events were analyzed using the {\tt Madanalysis5} package \cite{Araz:2020lnp}. 

We begin our analysis by demanding at least four charged leptons $N_{\ell} \geq 4$ in the final state. 
We ensure the isolation of all reconstructed objects by fixing $\Delta R=\sqrt{\Delta\phi^2 + \Delta\eta^2}\geq 0.4$ between any pair of particles, where $\Delta\phi$ and $\Delta\eta$ are their separations in the azimuthal angle and rapidity parameters, respectively. In addition, we have implemented the following basic acceptance cuts 

\begin{itemize}
    \item The charged leptons must satisfy the transverse momentum $p^{\ell}_{T} >$ 10 GeV, and rapidity $|\eta_{\ell}| \leq 2.5$ cuts. In addition, the invariant mass corresponding to the OSSF lepton pairs must satisfy $M_{\ell^{+}\ell^{-}} > 5$ GeV. 
    \item We also demand a veto on any b-jets with $p^{b}_{T} >$ 30 GeV and $|\eta_{b}| <$ 2.5. This will help us reject a large portion of the $t\bar{t}V,~t\bar{t}h, Vh$ background, as all these processes give rise to multiple $b$-quarks in the final state. However, a part of the BSM signal events from $A_{2}$ production gets affected due to this cut as the CP-odd scalar dominantly decays via $A_{2} \to Z'h_{1}$ with $h_1$ decaying to $b\bar{b}$. 
\end{itemize}
\begin{figure}[h!]
\centering
\includegraphics[width=0.46\textwidth]{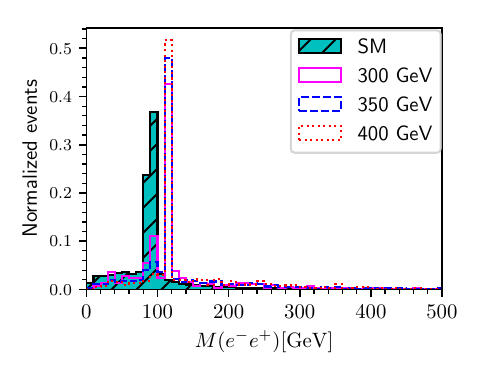} \hspace*{0.25in}
\includegraphics[width=0.46\textwidth]{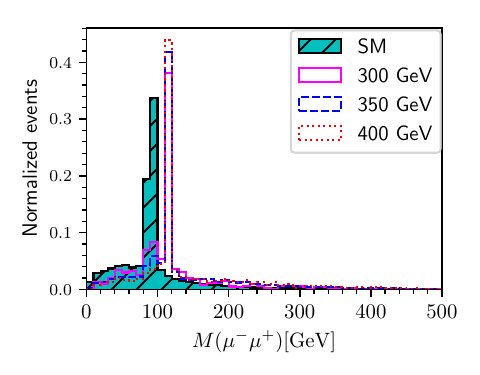}
\includegraphics[width=0.46\textwidth]{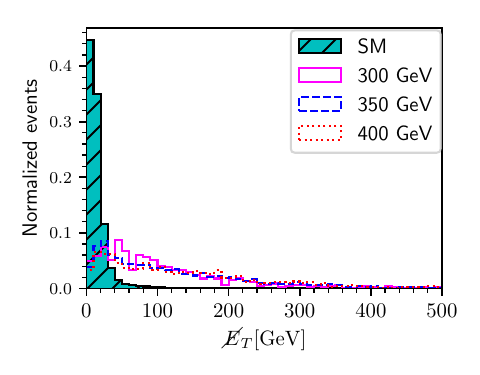} \hspace*{0.25in}
\includegraphics[width=0.46\textwidth]{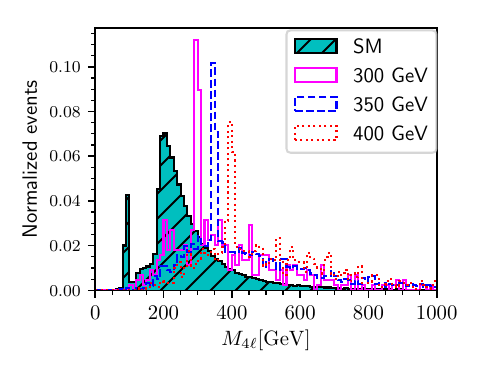}
\includegraphics[width=0.46\textwidth]{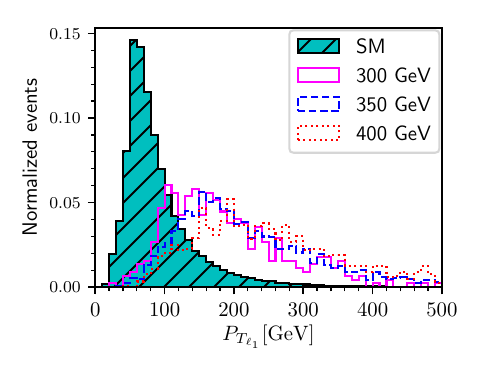} \hspace*{0.25in}
\includegraphics[width=0.46\textwidth]{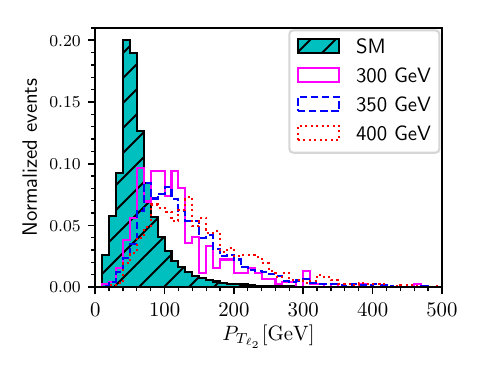}
\caption{Normalized distribution of the kinematic variables for $4 \ell + X$ final state at the $14$ TeV LHC.  } 
\label{fig:4lvarsB}
\end{figure}
For this study, we fix the light gauge boson mass at $M_{Z'} = 115$ GeV. Our goal here is to find suitable kinematic variables that can help us enhance the signal vs background ratio. In Fig.~\ref{fig:4lvarsB}, we present the normalized distribution of these variables. In the top row of Fig.~\ref{fig:4lvarsB}, we illustrate the invariant mass distribution of $e^{+}e^{-}$ and $\mu^{+}\mu^{-}$ pair for the signal and SM background. In the signal distribution, most of the OSSF lepton pairs come from the $Z'$ decay. As a result, one can see a sharp peak of around $115$ GeV for all the BPs. However, the neutral scalar $h_{2}$ can decay via $h_{2} \to Z Z'$ modes, and as a consequence, a part of the signal events peak around the $Z$ boson mass. In the case of SM, a dominant part of the OSSF lepton pairs are from the $Z$ decay. Therefore a cut to remove $Z$-peak in both $e^{+}e^{-}$ and $\mu^{+}\mu^{-}$ distributions would help us reduce a good fraction of the SM backgrounds, specifically for the case of $4\ell+jets$. 

Apart from the four leptons, the final state of the signal can also contain additional jets or missing energy as opposed to the pure $4\ell$ backgrounds. This motivates us to demand a missing transverse energy ($\mET$) cut for the signal while we inclusively keep all jets (including even hard jets with a minimum $p_T > 30$ GeV) into our signal events. To reduce the pure continuum $4\ell$ SM background we impose that the signal events must have a minimum 
$\mET >30$ GeV.
This helps us reject more than 90\% of the dominant SM background in the $4\ell+jets$. Our next goal is to find a suitable kinematic variable that further improves the signal efficiency against the SM background. In the bottom row of Fig.~\ref{fig:4lvarsB} we present the transverse momentum distribution of the leading ($p_{T}(\ell_{1})$) and sub-leading charged lepton ($p_{T}(\ell_{2})$), respectively. As pointed out earlier, most of the charged leptons originate from the decay of $Z'$ boson for the signal events, which is produced as a decay product of the heavy scalars. As a result, these leptons would be moderately boosted for all three benchmark scenarios. On the other hand, for the SM background, the charged leptons are produced due to the decay of $W$ and $Z$ bosons, and the corresponding distribution peaks around $p_{T}(\ell_{1}/\ell_{2}) \sim \frac{M_{W/Z}}{2}$. By imposing $p_{T}(\ell_{1}) > 90$ GeV and $p_{T}(\ell_{2}) > 60$ GeV cuts, we manage to reduce the SM background significantly while keeping the majority of the signal events. Finally, we have imposed $M_{4\ell} > 200$ GeV, which essentially captures the energy scale for the processes contributing to the signal events. As the effective scale of the BSM signal involves the production of heavier particles than the SM, a strong cut helps us improve the signal versus background ratio further.

\begin{table}[h]
	\centering
		\resizebox{15cm}{!}{
	\begin{tabular}{|c|c|c|c|c|c|c|c|c|c|}
	\cline{2-10}
	\multicolumn{1}{c|}{$\mathcal{L}=3000$ fb$^{-1}$} 
      & \multicolumn{6}{c|}{SM-background} &  \multicolumn{3}{c|}{Signal } \\  \hline	
                         Cuts  & ~~$4\ell$+jets~~ & ~~$VVV$~~   & ~~$t\bar t V$~~ &  ~~$W^{\pm}Z$~~ &~~$Vh$~~& ~~$t\bar t h$~~
                                  &  ~~{\tt BP1}~~ & ~~{\tt BP2}~~ & ~~{\tt BP3}~~ \\ \hline		
     $N_{\ell} \geq 4$, $M_{\ell^+\ell^-} > 5$~GeV, $N_{b} = 0$    & 44820.0 & 776.4 & 443.4  &  307.2 & 398.88  & 35.1 &92.37 &104.7&74.46\\ \hline	
     $M_{\ell^+\ell^-} < 80$~GeV or $M_{\ell^+\ell^-} >100$~GeV & 3864.9 & 9.6   &  21.24  &  82.59 & 22.71  & 19.32 & 38.7&56.67&48.3 \\  \hline	         		 
     $N_{j} > 0$    & 1399.2 & 6.45 & 20.31   &49.56 & 7.32  & 17.07 &35.34 &51.57&43.5\\  \hline	        
     $\mET >30$~GeV  &102.72 & 4.83 & 18.42 & 39.63 & 5.1  & 15.57  &29.34 &40.11&35.4\\  \hline	
     $p_{T_{\ell_1}} > 90$~GeV  &35.76 & 3.21& 10.89 & 13.2 & 1.5  & 7.98  &27.0 &38.46&34.59\\  \hline
     $p_{T_{\ell_2}} > 60$~GeV  &22.41 & 1.62 & 6.6 & 9.9 & 0.78  & 5.4  &24.99 &36.9&33.27\\  \hline
     $M_{4\ell} >200$~GeV &19.38 &1.62 & 6.6 & 3.3 & 0.66 & 5.13  & 24.18  &36.33&33.27\\  \hline
   \multicolumn{1}{|c|}{Total Events after cuts} 
      & \multicolumn{6}{c|}{36.6} & 24.18  &36.33 &33.27 \\  \cline{1-8}  \hline	         
      & \multicolumn{6}{c|}{Significance ($\mathcal{S}$)}  
      & 3.64 & 5.28  & 4.88 \\ \hline \hline
	\end{tabular}}
	\caption{ The cut-flow information on the $p \, p \rightarrow 4 \ell+X$ process for both the signal and 
	background along with the significances for {\tt BP1, BP2, BP3} at the 14 TeV LHC for 3000 fb$^{-1}$ integrated luminosity.}
	\label{tab:cutflow4l1}
\end{table}

We present the result of our analysis along with the respective cuts in Table.\ref{tab:cutflow4l1} for an integrated luminosity $\mathcal{L} = 3000 fb^{-1}$.  The signal significance $\mathcal{S}$ is calculated using the formula 
\begin{equation}
    \mathcal{S} = \sqrt{2\left[\left(s+b\right)\log\left(\frac{s+b}{b} -s\right)\right]}
\end{equation}
where $s$ and $b$ denote the signal and background events respectively. The significance corresponding to each of the benchmark scenarios is mentioned in the last row of Table.\ref{tab:cutflow4l1}. One can notice that the signal process corresponding to $\tt BP2$ shows maximum discovery potential as compared to $\tt BP1$ and $\tt BP3$. For $\tt BP3$, the scalar production cross section is the lowest but it still gives significance of around $5\sigma$ because of the larger branching probability to the $Z'$ as the BP has the maximum value of $g_x$. On the other hand for $\tt BP1$, the BSM scalar production is the maximum but it yields a smaller branching probability to $Z'$ since $g_{x}= 0.3$ is the lowest value in all BP's. Note that a larger $g_x$ for {\tt BP1} would give a greater significance but also take it closer to exclusion from the existing ATLAS analysis discussed in the previous section.  We thus find that the inclusive multilepton signal for scalars in a 2HDM gains if a light $Z'$ that is very weakly coupled to the SM sector can contribute to its decay modes. The reach for the scalars would improve significantly and scalars with masses above 350 GeV will be discoverable at the very high luminosity option of the LHC.

\section{Summary and conclusion}
\label{sec:summary}
The 2HDM scalar sector is a widely studied BSM scenario with interesting phenomenological signals. The minimal models of 2HDM where the SM is only extended by another Higgs doublet have very limited channels of discovery at the LHC. In this work, we consider a 2HDM in a model with $U(1)$ gauge extension of the SM that is responsible for light neutrino masses and a possible DM candidate. The scalar sector of the model has a doublet scalar and two SM singlet scalars that are charged under the new gauge symmetry. All the SM fields are neutral under this additional $U(1)_X$ symmetry. These new scalars break the new gauge symmetry spontaneously resulting in a massive gauge boson $Z'$ in the particle spectrum. Additional SM singlet fermions in the model are responsible for the SM neutrinos becoming massive via the popular inverse seesaw mechanism. The $Z'$ can interact with the SM fields via the GKM and the $Z-Z'$ mixing angle $\theta'$ that is restricted to values $\leq 10^{-3}$ for light $Z'$ with mass $\leq 150$ GeV, making its direct production at colliders very difficult. 

We revisit the signals for the scalars of the 2HDM in this model where a new channel opens up into the light $Z'$. We consider its production at LHC from the decay of the scalars from the second doublet as they are charged under the $U(1)_X$ symmetry.  These scalars, produced at the LHC, decay into boson pairs via channels such as $(H^{\pm} \to W^\pm Z', h_2 \to VV, (V = W^\pm, Z, Z'), A_2 \to h_1 Z'(Z))$. The subsequent decay of the unstable particles can produce multilepton final states that would extend the reach of the scalars beyond current bounds. We choose the most sensitive of all the final states, i.e. the $4\ell+X$ signal, and analyze the discovery prospects at the LHC. We find that the multilepton searches by the CMS and ATLAS collaboration constrain our parameter space. We evaluate the bounds on our model parameter space using the ATLAS search of four lepton final states using {\tt Rivet}.  We then consider some benchmark points with $M_{Z'}=115$ GeV for different values of the scalar masses and perform a cut-based analysis to show the signal sensitivity for the 2HDM scalars in the $4\ell+X$ channel. We find that the presence of a light $Z'$ allowed by current experimental bounds improves the signal sensitivity and has discovery potential for scalars with mass greater than 350 GeV with $3 \rm{ab^{-1}}$ integrated luminosity. This reach is dependent on the $U(1)_X$ gauge coupling ($g_x$) which controls the decay probability of the scalars into the $Z'$. A large value of this coupling will improve the reach further as the decay branching probability of the scalars to the $Z'$ increases with $g_x$. 
\begin{acknowledgements}
The author would like to acknowledge the support from the Department of Atomic Energy (DAE), India, for the Regional Centre for Accelerator-based Particle Physics (RECAPP), Harish Chandra Research Institute. AS also acknowledges the financial support from the SERB-National Postdoctoral fellowship (Ref No: PDF/2023/002572). 
\end{acknowledgements}
\appendix
\section{Feynman rules}\label{app:feynrule}
We describe below the $Z'$ couplings with SM fermions. Let $s_W \equiv \sin\theta_W$ and $c_W \equiv \cos\theta_W$, where $\theta_W$ refers to the Weinberg angle. Similarly, we define $s_{\theta'} \equiv \sin\theta'$ and $c_{\theta'} \equiv \cos\theta'$, with $\theta'$ being the mixing angle between the $Z$ and $Z'$ bosons. The isospin is represented by $T_3$, and the electric charge of the fermions is denoted by $Q_f$. The projection operators are expressed as $P_{L/R} = \frac{1 \mp \gamma_5}{2}$.
\begin{wrapfigure}[6]{l}{0.15\textwidth}
	\includegraphics[width=0.15\textwidth]{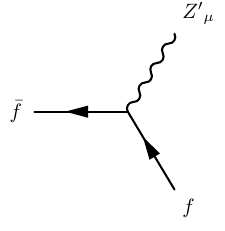}
\end{wrapfigure}
{\small\begin{eqnarray*} 
	i\left(\frac{e \, s_{\theta'}}{s_{W} c_{W}} \left(T_3 - Q_f s^2_{W}\right) + g_x' c_{\theta'}\left(T_3 - Q_f\right)\right)\!\gamma^\mu P_L - i\left(\frac{e \, s_{\theta'}}{s_{W} c_{W}} Q_f s^2_{W} + g_x' c_{\theta'}Q_f\right)\!\gamma^\mu P_R 
	\end{eqnarray*}}

\section{Decay width}\label{app:decaywidth}
\begin{flalign*} 
  & \Gamma_{Z'\to f \bar{f}}= \frac{M_{Z'} \sqrt{1 - \frac{4 m_f^2}{M_{Z'}^2}}}{12 \pi} \left( v_f^2 \left(1 + \frac{2 m_f^2}{M_{Z'}^2}\right) + a_f^2 \left(1 - \frac{4 m_f^2}{M_{Z'}^2}\right) \right) \,\, , && \\ \\
 & \Gamma_{H^+\to t \bar{b}} =  \frac{{3  y_t^2 \tan^2\beta}}{{16  \pi  m_{H^+}}} \left(m_{H^+}^2 - m_t^2 - m_b^2\right) \left(\sqrt{1 - 2 \frac{{m_b^2 + m_t^2}}{{m_{H^+}^2}} + \frac{{(m_t^2 - m_b^2)^2}}{{m_{H^+}^4}}}\right) \,\, ,&& \\ \\
 & \Gamma_{H^+\to W^+ Z'} = \frac{{(g_x  v_2 g_2)^2}}{{16  \pi  m_{H^+}}} \left(2 + \frac{{(m_{H^+}^2 - m_W^2 - m_{Z'}^2)^2}}{{4  m_{Z'}^2 m_W^2}}\right) \left(\sqrt{1 - 2 \frac{{m_W^2 + m_{Z'}^2}}{{m_{H^+}^2}} + \frac{{(m_W^2 - m_{Z'}^2)^2}}{{m_{H^+}^4}}}\right) \,\,\, ,&& \\
& \Gamma_{h_2\to Z' Z'} = \frac{{(g_{h_2Z'Z'})^2}}{{32  \pi  m_{h_2}}} \left(2 + \frac{{(m_{h_2}^2 - 2m_{Z'}^2)^2}}{{4  m_{Z'}^4}}\right) \left(\sqrt{1 -  \frac{{ 4m_{Z'}^2}}{{m_{H^+}^2}} }\right) \,\,\, ,&& \\
 & \Gamma_{h_2\to Z Z'} = \frac{{(g_{h_2ZZ'})^2}}{{16  \pi  m_{h_2}}} \left(2 + \frac{{(m_{h_2}^2 - m_Z^2 - m_{Z'}^2)^2}}{{4  m_{Z'}^2 m_Z^2}}\right) \left(\sqrt{1 - 2 \frac{{m_Z^2 + m_{Z'}^2}}{{m_{h_2}^2}} + \frac{{(m_Z^2 - m_{Z'}^2)^2}}{{m_{h_2}^4}}}\right) \,\,\, ,&& \\\\
 & \Gamma_{A_2\to h_1 Z'} = \frac{{(g_{A_2h_1Z'})^2 m^3_{A_2}}}{{16  \pi  }m_{Z'}^2}  \left(\sqrt{1 - 2 \frac{{m_{h_1}^2 + m_{Z'}^2}}{{m_{A_2}^2}} + \frac{{(m_{Z'}^2-m_{h_1}^2)^2}}{{m_{A_2}^4}}}\right)^3 \,\,\, .&& \\
\end{flalign*}
The expressions for the couplings $g_{h_2Z'Z'}$, $g_{h_2ZZ'}$ and $g_{A_2h_1Z'}$ can be found in the Table.\ref{tab:feynrules}.

%
\end{document}